\documentclass[11pt]{article}

\usepackage[T1]{fontenc}
\usepackage[utf8]{inputenc}

\newcommand{\RR}{\mathrm{R}}
\newcommand{\LL}{\mathrm{L}}
\newcommand*{\Lax}{\mathcal{L}}

 \newcommand{\wt}[1]{\widetilde{#1}}

\usepackage[a4paper,top=30.6mm,bottom=38.6mm,left=34mm,right=34mm,footskip=1.3cm]{geometry}

\usepackage[nofancyfonts]{Packages}
\usepackage{bbm}
\allowdisplaybreaks[2]

\acrodef{ybd}[\textsc{yb} deformation]{Yang--Baxter deformation}
\acrodef{nc}[\textsc{nc}]{non-commutativity}
\acrodef{sym}[\textsc{sym}]{super Yang--Mills}
\acrodef{himr}[\textsc{himr}  background]{Hashimoto--Itzhaki--Maldacena--Russo background~\cite{hep-th/9907166,hep-th/9908134}}
\acrodef{cybe}[\textsc{cybe}]{classical Yang--Baxter equation}
\acrodef{kl}[\textsc{kl}]{Kosmann--Lie}
\acrodef{sw}[\textsc{sw}]{Seiberg--Witten}
\acrodef{susy}[\textsc{susy}]{supersymmetry}
\acrodef{wznw}[\textsc{wznw}]{Wess--Zumino--Novikov--Witten}
\acrodef{eom}[\textsc{eom}]{equations of motion}
\acrodef{dof}[\textsc{dof}]{degrees of freedom}

\newcommand*{\setR}{\ensuremath{\mathbb{R}}}
\newcommand*{\setZ}{\ensuremath{\mathbb{Z}}}
\newcommand*{\del}{\mathop{\mathrm{{}\partial}}\mathopen{}}

\usepackage{hyperref}

\usepackage{cite}

\usepackage{fancyhdr}
\newcommand{\preprint}[1]{%
  \fancypagestyle{empty}{%
    \fancyhf{} %
    \fancyheadoffset[R]{\marginparsep+\marginparwidth}
    \fancyhead[R]{{#1}}
    \renewcommand{\headrulewidth}{0pt}
    \renewcommand{\footrulewidth}{0pt}
  }
}

\newcommand*{\AdS}[1]{\ensuremath{\mathrm{AdS}_{#1}}}

\newcounter{example}

\begin{document}

\numberwithin{equation}{section}

\preprint{KUNS-2767}

\begin{center}
  {\vspace*{1.5cm}\Huge $O(d,d)$ transformations preserve\\[5pt]
    classical integrability}
\vspace*{1.5cm}\\
{\large 
Domenico Orlando$^{\sharp,\flat}$\,, 
Susanne Reffert$^{\flat}$\,,
Yuta Sekiguchi$^{\flat}$\,,
and Kentaroh~Yoshida$^{\ast}$
} 
\end{center}
\vspace*{0.3cm}
\begin{center}
  $^{\sharp}$\emph{ INFN, sezione di Torino and Arnold--Regge Center\\[1pt]
  via Pietro Giuria 1, 10125 Torino, Italy}
\vspace*{0.25cm}\\ 
$^{\flat}$\emph{Institute for Theoretical Physics, 
Albert Einstein Center for Fundamental Physics,\\[1pt] 
University of Bern, Sidlerstrasse 5, CH-3012 Bern, 
Switzerland 
}
\vspace*{0.25cm}\\ 
$^{\ast}$\emph{Department of Physics, Kyoto University, \\[1pt]
Kyoto 606--8502, Japan} 
\end{center}
\vspace{0.3cm}

\vfill

\begin{abstract}
  \noindent
 In this note, we study the action of $O(d,d)$ transformations on the integrable structure of two-dimensional non-linear sigma models via the doubled formalism.
  We construct the Lax pairs associated with the $O(d,d)$-transformed model and find that they are in general non-local because they depend on the winding modes.
    We conclude that every $O(d,d;\setR)$ deformation preserves integrability.
  As an application we compute the Lax pairs for continuous families of deformations, such as $J\bar{J}$ marginal deformations and TsT transformations of the three-sphere with $H$-flux.

\end{abstract}
\vfill

\setcounter{footnote}{0}
\setcounter{page}{0}
\thispagestyle{empty}

\newpage
\tableofcontents
\newpage

\section{Introduction}%
\label{sec:introduction}

This note presents a synthesis of ideas which separately have been subject to intense study. On the one hand, we have non-linear sigma models and their deformations, such as marginal current-current deformations preserving conformality. On the other hand we have the integrability which allows the use of extremely powerful computational techniques and the study of integrable deformations of sigma models which preserve this property. We observe that $O(d,d)$ transformations give rise to examples in both groups of deformations. Using the $O(d,d)$ invariant doubled formalism, we show on general grounds that a sigma model which is classically integrable remains so under any deformation generated by an $O(d,d)$ transformation.

\bigskip
Two-dimensional non-linear sigma models form the basis of the world-sheet description of strings propagating in target space.
They encode the massless excitations of the string, namely the metric $G_{\mu\nu}$, the anti-symmetric Kalb--Ramond field $B_{\mu\nu}$, and the dilaton.
The field equations which determine the dynamics of these fields are associated to the beta function of the string sigma model:
$G$, $B$ and $\Phi$ will solve the Einstein field equations if the corresponding sigma model is conformal.   
Understanding the moduli space of such two-dimensional conformal field theories is crucial for our understanding of string theory.
It is therefore interesting to study the effect of marginal deformations which preserve conformal invariance.
One of the traditional approaches is to deform a given theory using a bilinear of left and right-moving conserved currents~\cite{Hassan:1992gi,Henningson:1992rn,Kiritsis:1993ju,Giveon:1993ph,Forste:1994wp,Israel:2003ry,Forste:2003km} associated to the isometry group of the target space.
Recently, it has been suggested that this type of construction is related to the gravity duals of the $T\bar{T}$ and \(J\bar{T}\) deformation~\cite{McGough:2016lol,Apolo:2018qpq,Giveon:2017nie,Chakraborty:2018vja,Borsato:2018spz,Araujo:2018rho,Chakraborty:2019mdf}.
Such current-current deformations can be understood as a rotation within the  $O(d,d;\setR)$ group that generalizes the \(O(d,d;\setZ)\) symmetry of string theory\cite{Chaudhuri:1988qb,Giveon:1988tt,Duff:1989tf,Giveon:1991jj,Gasperini:1991qy,Maharana:1992my,Israel:2004vv,Israel:2004cd,Detournay:2005fz,Orlando:2006cc,Rennecke:2014sca}.
For reviews, see~\cite{Giveon:1994fu,Alvarez:1994dn,Maharana:2013uvy}.
Also, T-duality, one of the characteristic features of string theory, is part of this $O(d,d)$ group. 
Its essence lies in the interplay between the momenta and winding modes of closed strings. 
At the level of the sigma model, T-duality is a consequence of gauging the commuting isometry group on the target space~\cite{Buscher:1987sk,Buscher:1987qj}.%

\medskip

The duality-invariant formulation of string theory has a long history~\cite{Duff:1989tf,Tseytlin:1990nb,Tseytlin:1990va,Maharana:1992my}.
The idea is based on the introduction of winding coordinates.
A sigma model with a $T^{d}$ fiber is naturally extended to an enlarged sigma model endowed with a $T^{2d}$ fibration containing also the winding coordinates subject to a consistency constraint~\cite{Hull:2004in,Hull:2006va,Hull:2006qs,Hull:2007jy,Hull:2009sg}.
The resulting double sigma model action is \emph{manifestly} invariant under $O(d,d; \setZ)$ transformations.
The doubled formalism 
is particularly effective to describe so-called non-geometric backgrounds~\cite{Dabholkar:2002sy,Dabholkar:2005ve}, where the transition functions between local patches involve a T-duality transformation, generalizing the notion of geometry.\footnote{For a recent review of non-geometric backgrounds, see~\cite{Plauschinn:2018wbo}.} 

\bigskip
The relationship between integrability and T-duality is by now well understood.
The existence of a Lax pair, giving rise to an infinite number of (non-local) conserved charges, is a sufficient condition for classical integrability. The construction of non-local charges is given in~\cite{Luscher:1977uq,Luscher:1977rq,Brezin:1979am}.
In~\cite{Ricci:2007eq} it was shown that starting from a model with a known Lax pair, it is possible to construct the Lax pair for any T-dual system. Related works are~\cite{Hatsuda:2006ts,Kluson:2008nn,Orlando:2010yh,Orlando:2012hu}.
In general, the corresponding conserved charges are non-local. %
In the context of integrability~\cite{Beisert:2010jr} of the AdS/CFT correspondence~\cite{Maldacena:1997re}, the non-local charges of the (T-dual of) $AdS_{5}\times S^{5}$ strings were studied in~\cite{Bena:2003wd,Alday:2003zb,Hatsuda:2004it,Beisert:2008iq,Beisert:2009cs,Hatsuda:2011mt}.
T-duality also plays a role in the study of Yang--Baxter deformations~\cite{Klimcik:2002zj,Klimcik:2008eq,Delduc:2013qra,Kawaguchi:2014qwa,Klimcik:2014bta,Klimcik:2015gba,Klimcik:2016rov,Klimcik:2017ken}, a technique which has led to the construction of integrable deformations of \AdS{} backgrounds as well as Minkowski spacetime.
Some of these backgrounds can be understood in terms of TsT transformations~\cite{Hashimoto:1999ut,Maldacena:1999mh,Lunin:2005jy,Alday:2005ww,Frolov:2005dj,Matsumoto:2014ubv,Matsumoto:2014nra,Crichigno:2014ipa,Matsumoto:2014gwa,Matsumoto:2015ypa,Matsumoto:2015uja,vanTongeren:2015soa,Osten:2016dvf,Delduc:2017fib}, non-Abelian T-dualities~\cite{Hoare:2016wsk,Borsato:2016pas,Sakamoto:2016ppx,Borsato:2018idb,Lust:2018jsx}, or generalized T-dualities~\cite{Arutyunov:2015mqj,Orlando:2016qqu}.
The latter gives rise to solutions of the generalized supergravity equations~\cite{Arutyunov:2015mqj,Wulff:2016tju}, in which an extra vector field gives rise to non-geometric $Q$-fluxes~\cite{Fernandez-Melgarejo:2017oyu} (For a physical interpretation of the extra vector, see also~\cite{Araujo:2017jkb,Araujo:2017enj,Araujo:2017jap}).
Generalized supergravity equations have been also studied in the T-duality invariant framework~\cite{Sakatani:2016fvh,Sakamoto:2017wor,Catal-Ozer:2019tmm}.

\medskip

From the point of view of string theory, T-duality is part of a larger \(O(d,d; \setZ )\) symmetry that in turn can be extended to the group \(O(d,d; \setR)\).
Note that transformations in \(O(d,d; \setR) \setminus O(d,d;\setZ)\) are not symmetries. We will use them instead as solution-generating operations.
In this paper we study the interplay between the action of the $O(d,d)$ group and classical integrability of sigma models using the doubled formalism.
Extending the argument in~\cite{Ricci:2007eq}, we present a systematic approach to construct $O(d,d)$-deformed Lax pairs.
We start with a two-dimensional sigma model \(\mathcal{S}\) on a manifold with isometry group \(G\) of rank $d$.
We pick a maximal torus \(T^d \subset G\) as a fiber and we choose a coordinate system such that \(T^d \) is generated by the Killing vectors \(\ev{\del_{x^1}, \dots , \del_{x^n}} \).
Then we double the torus and introduce doubled adapted coordinates \(\{ x^1, \dots x^d, \wt{x}_1, \dots, \wt{x}_d \}\) on which \(O(d,d)\) acts linearly.
What we obtain in this way is a natural action of the group \(O(d,d;\setZ)\), which can be extended to \(O(d,d; \setR)\): for each element \(g \in O(d,d; \setR)\) we obtain a new (generically inequivalent) sigma model \(\mathcal{S}'\), which in general has only isometry group \(T^d\), and not the original group $G$.
Using Noether's procedure we write down the conserved currents \(J \in \mathfrak{g}\) for the initial model.
If these currents are conserved and flat, they can be used to introduce a Lax pair \(\Lax\) that guarantees the classical integrability of the initial model.
The \(O(d,d)\) transformation \(g\) maps the Lax pair into a new pair \( \Lax'\) for \(\mathcal{S}'\).
These currents \emph{do not} in general correspond to isometries of \(\mathcal{S}'\), and so \emph{do not} stem from Noether's construction. They are \emph{non-local}.
Nonetheless, they are still a one-parameter family of flat currents which is enough to guarantee the classical integrability of the deformed model \(\mathcal{S}'\).
The \(O(d,d)\) transformation maps the momenta of \(\mathcal{S}\) generically into momenta and winding modes of \(\mathcal{S}'\).
The non-locality of the currents is a consequence of the fact that the Lax pair \(\Lax'\) is constructed using these windings.
Having a system that has \emph{more currents than isometries}, the extra currents being realized in terms of winding modes, should hardly be surprising.
The simplest and best-known example of this effect is the compact boson at the self-dual radius that has an \(SU(2) \times SU(2)\) symmetry as opposed to the geometrical \(U(1) \times U(1)\). The extra symmetry is naturally interpreted in terms of winding modes~\cite{Polchinski:1998rq}.

\bigskip

The plan of this paper is as follows.
In Section~\ref{sec:Odd-general} we introduce the basics of the doubled formalism and the group $O(d,d)$.
In Section~\ref{sec:Odd-currents} we remind ourselves of the construction of the Lax pairs and conserved charges in classical integrability and how the Lax pair transforms under $O(d,d)$.
In Section~\ref{sec:O11} we put the general formalism into practice using the simple example of $O(1,1)$-transforming the sigma model on the two-sphere.
In Section~\ref{sec:O22} we study the \ac{wznw} model on the three-sphere. Specific $O(2,2)$ elements give rise to $J\bar J$ deformations, TsT transformation and double T-duality of the original sigma model.
For all these cases, we explicitly construct the transformed Lax pairs, thus making the classical integrability of the deformed models manifest.
In Section~\ref{sec:conclusions} we conclude and present a number of future directions in which our work could be extended.

\section{Non-linear sigma models and $O(d,d)$ transformations}%
\label{sec:Odd-general}

Let us first review the basics of the doubled formalism and $O(d,d)$ transformations.

We consider a geometric string background, where the $D$-dimensional target manifold is equipped with a metric $G$ and the closed three-form $H$-flux. $H$ has locally a two-form potential $B$. We do not consider the dilaton in this paper. Defining a set of local coordinates $X^{\hat \imath}$, $ \hat \imath=1, \dots, D$, the string sigma model action is given by
\begin{equation}
  \label{eq:standard-sigma}
  S[G, B] = \frac{1}{2}\int G_{\hat \imath \hat \jmath}(X) \dd X^{\hat \imath} \wedge \star \dd X^{\hat \jmath} +  B_{\hat \imath \hat \jmath}(X) \dd X^{\hat \imath} \wedge \dd X^{\hat \jmath}\,,
\end{equation}
where the Hodge duality on the world-sheet satisfies $\star^{2} = 1$.
We assume that the manifold of  interest has Euclidean signature. 

We will use the doubled formalism, which is motivated by the search for a back\-ground-independent formulation of string theory and is manifestly invariant under $O(d,d)$ transformations, where $d$ denotes the dimension of the maximal torus $T^{d}$. 
From the string sigma model, we read off the metric $G_{ij}$ and the $B$-field $B_{ij}$ in the isometric directions $i,j=1,\dots,d$ and package them into the so-called generalized metric
\begin{align}
{\mathcal{H}(G, B )}_{\hat I \hat J} &= 
	\pmqty{
	G - BG^{-1}B & BG^{-1}\\
	- G^{-1} B & G^{-1}
 	}_{\hat I \hat J} , &  \hat I, \hat J &=1,\dots,2d\,.
\end{align}
It is not hard to see that the generalized metric satisfies 
\begin{equation}
\mathcal{H}^{t} L \mathcal{H} = L\,,
\end{equation}
where the indefinite matrix $L$ is given by
\begin{equation}
\label{eq:Odd-metric}
L = \pmqty{
	0 & \mathbbm{1} \\
	\mathbbm{1} & 0
	}.
\end{equation}
The $2d \times 2d$ generalized metric can be considered as the curved metric of the $2d$-dimensional doubled space. 

Our target space manifold has a group of isometries \(G\).
Let us focus on a submanifold $M$ of the target space, on which the maximal torus \(T^d \subset G\), \(d \le D\) of the full isometry group acts freely.
In other words we separate the coordinates \(X^{\hat \imath}\) into those that describe \(M\) (we will call them \(X^i\)) and those that describe the base (\(Y\)).
In the doubled formalism we extend \(M\) to a doubled manifold, whose local patches are formed by a patch of $M$ and a patch of the T-dual $\wt{M}$.
We denote a set of local coordinates on a patch of the doubled manifold by $\mathbb{X}^{I}, I=1, \dots ,2d$. It consists of the doublet of local coordinates on  $M$ and $\wt{M}$: 
\begin{align}
\mathbb{X}^{I} &= 
	\pmqty{
	X^{i}\\
	\wt{X}_{i}
	}, & i&=1,\dots,d\,.
\end{align}
In the following we will always choose the polarization such that the first half of the components of the doubled coordinates are the ``physical'' ones.

The action in the doubled formalism is given by~\cite{Hull:2004in,Hull:2006va,Hull:2006qs,Hull:2007jy,Hull:2009sg}
\begin{equation}
\label{eq:doubled-sigma}
S_{d} =  \int \frac{1}{2} \mathcal{H}_{IJ} \dd \mathbb{X}^{I} \wedge \star \dd \mathbb{X}^{J} + \dd \mathbb{X}^{I} \wedge \star \mathcal{J}_{I}(Y) + \mathcal{L}(Y)\,,
\end{equation}
where $\mathcal{J}_{I}$ is the source term dependent only on $Y$ and $\mathcal{L}(Y)$ is the Lagrangian density dependent only on $Y$. The action is manifestly invariant under $O(d,d; \setZ)$ transformations since for $g \in O(d,d; \setZ)$ the action transforms according to
\begin{align}%
  \label{eq:coord-redef}
\mathcal{H} &\rightarrow g^{t} \mathcal{H} g\,, & \dd \mathbb{X} &\rightarrow g^{-1} \dd \mathbb{X}\,, & \mathcal{J} &\rightarrow g^{t} \mathcal{J}\,.
\end{align}
From these transformation rules we can reconstruct the deformed string sigma model with fields $G'$ and $B'$ in terms of the new local coordinates $X'^{i}$ (see Sec.~\ref{sec:deformed-metric-and-B}).
As we will see later, the field equations for $X$ and $Y$ are equivalent, under the self-duality constraint in Eq.~\eqref{eq:self-duality},  to those coming from the standard sigma model action. 

\medskip

Once the action of \(O(d,d;\setZ)\) is defined in this way, we can generalize it to \(O(d,d; \setR)\). In the following, we use the action of this latter group as a solution generating technique.
The $O(d,d; \setR)$ elements include real continuous parameters, which act generically as  deformation parameters for the starting model.

\subsection{The group $O(d,d)$}%
\label{sec:Odd-group}

We next present the properties of the group $O(d,d)$ in detail.  We write $g \in O(d,d)$ as  
\begin{equation}
\label{eq:Odd-element}
g = 
	\pmqty{
	\alpha  & \beta\\
	\gamma & \delta 
	}\,,
\end{equation}
where $\alpha$, $\beta$, $\gamma$ and $\delta$ are $d\times d$ matrices. Their index structures are given by
\begin{equation}
\alpha^{i}{}_{j}\,,\qquad \beta^{ij}\,,\qquad \gamma_{ij}\,,\qquad \delta_{i}{}^{j}\,,\qquad i,j=1, \dots ,d\,.
\end{equation}
The matrix \(g\) leaves the indefinite metric~\eqref{eq:Odd-metric} invariant,
\begin{align}
g^{t} L \,g &= L\,,
\end{align}
where the block matrices satisfy
\begin{equation}
\begin{cases}
\alpha^{t} \gamma + \gamma^{t} \alpha &= 0\\
\beta^{t} \delta + \delta^{t} \beta &=0\\
\alpha^{t} \delta + \gamma^{t} \beta &= \mathbbm{1}
\end{cases}
\,,\qquad\qquad
\begin{cases}
\alpha \beta^{t} + \beta \alpha^{t} &= 0\\
\gamma \delta^{t} + \delta \gamma^{t} &=0\\
\alpha \delta^{t} + \beta \gamma^{t} &=\mathbbm{1}
\end{cases}
\,.
\end{equation}
The inverse of $g$ is given by
\begin{equation}
\label{eq:Odd-inverse}
g^{-1} = \pmqty{ \delta^{t} & \beta^{t}\\ \gamma^{t} & \alpha^{t}}.
\end{equation}
The elements of this group are generated by the following elements. 

\paragraph{Diffeomorphisms.}
Given any invertible $d\times d$ matrix $A$, an element of the group of diffeomorphisms is parametrized by
\begin{equation}
\label{eq:diffeo}
g_{A} = 
		\pmqty{
		A & 0 \\
		0 & \left(A^{-1}\right)^{t}
		}\,,
\end{equation}
corresponding to a general coordinate transformations of the metric and $B$-field.
In terms of the generalized metric,
\begin{equation}
g_{A}^{t}\mathcal{H}(G, B)\,g_{A} = \mathcal{H}(A^{t}GA, A^{t}BA)\,.
\end{equation}

\paragraph{$B$-shift.}
The gauge symmetry of the $B$-field is also naturally encoded in the doubled formalism. For an exact two-form $\dd \Lambda$, the matrix for a shift of the $B$-field is given by
\begin{equation}
\label{eq:B-shift}
g_{B} = \pmqty{ \mathbbm{1} & 0 \\ - \dd \Lambda & \mathbbm{1} }\,,
\end{equation}
which leads to
\begin{equation}
g_{B}^{t}\mathcal{H}(G, B)g_{B} = \mathcal{H}(G, B + \dd \Lambda)\,.
\end{equation}

\paragraph{$\beta$-transformation.}
The conjugate to the $B$-shift is the so-called $\beta$-transformation. It is encoded by the bi-vector $\beta$ as
\begin{equation}
\label{eq:beta-shift}
g_{\beta} = \pmqty{ \mathbbm{1} & - \beta \\ 0 & \mathbbm{1}}\,.
\end{equation}
The frame after $\beta$-transformations is known as a non-geometric frame. The bi-vector corresponds to an antisymmetric tensor obtained by the Seiberg-Witten map~\cite{Duff:1989tf,Seiberg:1999vs}. The supergravity framework based on this frame is called $\beta$-supergravity~\cite{Andriot:2013xca}. 

\paragraph{$T$-duality.}
Finally, let us look at the matrix for the Abelian $T$-duality. We denote by $E_{k}$ the $d\times d$-matrix with 1 in the $(k,k)$-entry and 0 everywhere else. Then the matrix for T-duality along the $k$-th direction is given by
\begin{equation}
g_{T_{k}} = \pmqty{ \mathbbm{1} - E_{k} & E_{k} \\ E_{k} & \mathbbm{1}- E_{k}}\,.
\end{equation}
The metric and $B$-field transform according to the standard Buscher rules~\cite{Buscher:1987qj,Buscher:1987sk}.

\medskip

A general $O(d,d)$ element can be decomposed as
\begin{equation}
g = (g_{+})^{\eta_{+}}(g_{-})^{\eta_{-}}\prod_{i=1}^{n} g_{\beta_{i}}g_{B_{i}}g_{A_{i}}\,,
\end{equation}
where $\eta_{\pm} \in \{0,1\}$ and $g_{\pm}$ represents the quotient group $O(d,d)/ O(d,d)_{0}$ by the identity component $O(d,d)_{0}$:
\begin{equation}
g_{\pm} = \pmqty{ \mathbbm{1} - E_{1} & \pm E_{1} \\ \pm E_{1} & \mathbbm{1} - E_{1}}\,.
\end{equation}

\subsection{Deformed metric and $B$-field}%
\label{sec:deformed-metric-and-B}

Now we want to express the redefined metric and $B$-field after a transformation in terms of the fields in the original frame. We use an $O(d,d)$ element $g$ of the form Eq.~\eqref{eq:Odd-element} to rotate the generalized metric~\cite{Blumenhagen:2013aia}. 

The bottom right block gives the inverse of the redefined metric $G'$,
\begin{equation}
  (G')^{-1} = \left[ \delta + (G-B) \beta \right]^{t} G^{-1}\left[ \delta + (G-B)\beta\right]\,.
\end{equation}
The redefined metric is thus written as 
\begin{align}
\label{eq:redefined-metric}
  G' &= \rho^{-1}_{1} G (\rho^{-1}_{1})^t, & \text{with }\quad \rho_{1} &=  \delta + (G-B) \beta .
\end{align}
Next, we look at the top right block of the rotated generalized metric. It reads
\begin{equation}
  B'(G')^{-1} = -1 + \left[ \gamma + (G-B)\alpha\right]^{t} G^{-1}\rho_{1}\,.
\end{equation}
Multiplying by $G'$ from the right, we find
\begin{align}
\label{eq:redefined-B-field}
B' &= \rho_{1}^{-1}\left[ \rho_{1}\rho_{2}^{t} - G \right](\rho_{1}^{-1})^t,  & \text{with }\quad \rho_{2} &= \gamma + (G - B) \alpha.
\end{align}

\subsection{Self-duality constraint}

In order to get the right number of (physical) \ac{dof}, we need to impose an extra constraint on the doubled variables.

We start from a set of the pull-backs of $\dd \mathbb{X}^{I}$ which contain both physical and winding momenta:
\begin{align}
\dd \mathbb{X}^{I} &=  
\pmqty{
\dd X^{i}\\
\dd \wt{X}_{i}
}, & i&=1,\dots,d. 
\end{align}  
We now impose the \emph{self-duality constraint}
\begin{equation}
\label{eq:self-duality}
\dd \mathbb{X}^{I} = L^{IJ}\mathcal{H}_{JK} \star  \dd \mathbb{X}^{K},
\end{equation}
where the Hodge dual on the world-sheet satisfies $\star^{2} = 1$.
As long as this constraint holds, the \ac{eom} of the doubled sigma model are always satisfied~\cite{Hull:2004in,Hull:2006va,Hull:2006qs,Hull:2007jy,Hull:2009sg}.

\bigskip

In the following we will use the self-duality constraint to relate the physical coordinates of a model and of its dual.
The first component of the constraint can be rewritten as
\begin{equation}
  \label{eq:winding-current}
  \dd \wt{X}_{i} = \star \left( G_{ij} \dd X^{j} + B_{ij} \star \dd X^{j} \right) = \star J_{i} ,
\end{equation}
where the $J_{i}$ are the Noether currents associated to the freely acting $U(1)^{d}$ isometries along the Killing vectors $k_i= \partial_{X^{i}}$.
According to Eq.~\eqref{eq:winding-current}, the differentials of the winding coordinates $\wt{X}_{i}$ are interpreted as the Hodge duals of Noether currents $J_{i}$.
The fact that $\dd \wt{X}_{i}$ is exact implies the conservation of the currents $J_{i}$:
\begin{equation}
\dd^{2}\wt{X}_{i} = 0 = \dd \star\, J_{i} ,
\end{equation}
which plays the role of an on-shell condition for the dual model.

Under $g \in O(d,d)$, the $\dd \mathbb{X}$ are related to $\dd \mathbb{X}'$ as in Eq.~\eqref{eq:coord-redef} via
\begin{equation}
\label{eq:field-redef}
\dd \mathbb{X}^{I} = \tensor{g}{^I_J} \dd{\mathbb{X}'^{J}}\,.
\end{equation}
The differential of the physical coordinates in the original frame is given by
\begin{equation}
  \label{eq:X-and-Xp-and-Jp}
  \begin{aligned}
\dd X^{i} &=  \tensor{\alpha}{^{i}_{j}} \dd{X'^{j}} + \beta^{ij} \dd{\wt{X}'_{j}} = \tensor{\alpha}{^{i}_{j}} \dd{X'^{j}} + \beta^{ik} \star J'_{k}\,,\\
\dd{\wt{X}_{i}} &= \star J_{i} = \gamma_{ij} \dd{X'^{j}} + \tensor{\delta}{_{i}^{k}} \star J'_{k}\,.
\end{aligned}
\end{equation}
Their inverse is given by
\begin{equation}
  \begin{aligned}
    \label{eq:Xp-and-X-and-J}
    \dd X'^{i} &= \tensor{(\delta^{t})}{^{i}_{j}} \dd X^{j} + \left(\beta^{t}\right)^{ik} \star J_{k}\,,\\
    \dd \wt{X}'_{i} &= \star J'_{i} = (\gamma^{t})_{ij} \dd X^{j} + \tensor{(\alpha^{t})}{_{i}^{k}} \star J_{k}.
  \end{aligned}
\end{equation}
Using expression~\eqref{eq:winding-current}, we can deduce a relation between the physical coordinates of the original and of the dual model:
\begin{equation}
\label{eq:duality-map}
\begin{aligned}
\dd X^{i} &= \beta^{ij} G'_{jk} \star \dd X'^{k} + \tensor{(\alpha + \beta B')}{^{i}_{j}}\dd X'^{j} \,,\end{aligned}
\end{equation}
where the primed metric and $B$-field are given in Eq.~\eqref{eq:redefined-metric} and Eq.~\eqref{eq:redefined-B-field}.
This relation is the so-called \emph{$O(d,d)$-duality map} derived in~\cite{Rennecke:2014sca}.
Note that this equation describes the action of \(O(d,d)\) on the differentials \(\dd{X}\). It can be naturally extended to \(O(d,d;\setR)\) even though we have introduced it on the double torus where only \(O(d,d;\setZ)\) is a symmetry.
The  map is non-local in the sense that the Hodge duals of the conserved currents $J_{i}$ are involved.
We will use it in the following  to directly construct the $O(d,d)$-deformed Lax pairs.

\section{$O(d,d)$ transformed Lax pairs}%
\label{sec:Odd-currents}

We are now in the position to study how the property of classical integrability behaves under $O(d,d)$ transformations. To do so, we first remind ourselves of the construction of the Lax pair.

\subsection{Lax pairs and conserved charges}
If a system has a global symmetry \(G\) (which we assume to be a connected Lie group), we can use the Noether procedure to construct the corresponding conserved currents. We denote the Killing vectors associated to the isometries by $k_i$ and the conserved ($\dd \star J=0$) Noether currents by $J_{i}$.

These currents may additionally fulfill a flatness condition or Maurer--Cartan equation,
\begin{equation}
	\dd J + J \wedge J =0.
\end{equation}
This flatness is the underlying reason for the classical integrability of a model, as from a flat conserved current $J$ we can always construct the so-called Lax pair:
\begin{align}\label{eq:LaxConnection}
  \Lax_{\lambda} &= a(\lambda) J+ b(\lambda) \star J\,,
\end{align}
where $\lambda\in \setR$ is the spectral parameter. 
In order to preserve flatness,
\begin{equation}
  \label{eq:flatness}
  \dd \Lax_{\lambda} + \Lax_{\lambda} \wedge \Lax_{\lambda} = 0\,,
\end{equation}
we must set
\begin{align}\label{eq:LaxFlat}
  a(\lambda)&=\frac{1}{2}(1\pm \cosh(\lambda)) && \text{and} & b&=\frac{1}{2}\sinh(\lambda) .
\end{align}
The existence of the Lax pair assures the classical integrability of the model, as each of the $\Lax_{\lambda}$ gives rise to infinitely many non-local conserved charges.

The construction of these charges is by now standard~\cite{Luscher:1977rq}.
Consider the {Wilson line} $W(x, t; \lambda)$ %
defined as the path-ordered exponential of the Lax connection between $(x_0, t_0)$ and  $(x,t)$, 
\begin{equation}
  \label{eq:w}
  W(x, t| x_0, t_0; \lambda) = \mathop{\mathrm{P}}\left[  e^{ \int\displaylimits_{\mathcal{C}:(x_0,t_0) \to (x,t)} \Lax_{\lambda}}\right].
\end{equation}
Using \(W\) we can now define a one-parameter family of conserved charges (monodromy matrix):
\begin{equation}
  Q(t; \lambda) = W(+\infty, t\ |-\infty, t; \lambda) =
  \mathop{\mathrm{P}}\left[  \exp \left( \int_{-\infty}^\infty \Lax_{\lambda}(x) \dd x \right) \right].
\end{equation}
Since the Lax pair \(\Lax_{\lambda}\) is flat, if it also vanishes
at spatial infinity ($\Lax_{\lambda}(\pm \infty, t) = 0$), the
one-parameter charge $Q(t; \lambda)$ is conserved for any $\lambda$:
\begin{equation}\label{eq:cond}
  \dv{t} Q(t;\lambda ) = 0.
\end{equation}
Expanding around $\lambda = 0$,
\begin{equation}
  \label{eq:conserved-charges}
  Q(t; \lambda ) = 1 + \sum_{n=0}^\infty  \lambda^{n+1} Q^{(n)}(t) ,
\end{equation}
the condition in Eq.~\eqref{eq:cond} is equivalent to the conservation
of the infinite set of charges 
\begin{align}
  \dv{t} Q^{(n)}(t ) &= 0, & \forall \,n &= 0, 1, \dots.    
\end{align}
The construction of the infinitely many conserved charges is based only on the existence of the flat of $\Lax_{\lambda}$.
It can be performed for any model with a one-parameter family of Lax pairs $\Lax_{\lambda }$, which is then classically integrable regardless of the nature of the currents.
This is in particular also true for non-local currents which do not stem from a Noether construction. We will make use of this fact in the following.

\subsection{Flat connections and $O(d,d)$ transformations}

We have seen how the background data and physical coordinates transform under the action of $O(d,d)$. Suppose that the initial model is integrable in the sense of the existence of Lax pairs which satisfies the zero-curvature condition~\eqref{eq:flatness}. Now we want to show that the Lax pairs of the original model can be mapped to new Lax pairs under $O(d,d)$ transformations.

The $O(d,d)$ map in Eq.~\eqref{eq:duality-map} acts only on the differentials of adapted coordinates. Therefore, it is important to make sure that the Lax pairs depend on the adapted coordinates only through the derivatives. In other words we need to find a new set of flat currents that is manifestly invariant under the action of the maximal torus $T^{d}$. This can be realized by an appropriate gauge transformation under which the Lax pairs transform in the adjoint representation as
\begin{equation}
  \Lax_{\lambda} \rightarrow \hat{\Lax}_{\lambda} = h^{-1} \Lax_{\lambda} h + h^{-1} \dd h\,.
\end{equation}
where $h \in G$ with $G$ a symmetry group of the initial model. The gauged Lax pair has a vanishing curvature on-shell, as
\begin{equation}
  \dd \hat{\Lax}_{\lambda} + \hat{\Lax}_{\lambda} \wedge \hat{\Lax}_{\lambda} = h^{-1} \pqty{ \dd \Lax_{\lambda} + \Lax_{\lambda} \wedge \Lax_{\lambda} } h = 0,
\end{equation}
where the last equality is guaranteed by the equations of motion in the initial model.

Now that the gauged Lax pairs do not explicitly depend on the adapted coordinates, we apply the $O(d,d)$ map~\eqref{eq:duality-map} to find the $O(d,d)$-dual Lax pairs of the form
\begin{equation}
  \hat{\Lax}_{\lambda}(\dd X^{i}) \to  \Lax'_{\lambda}(\dd X'^{i}) = \hat{\Lax}_{\lambda}(\dd X^{i} \rightarrow \alpha^{i}{}_{j} \dd X'^{j} + \beta^{ik} \star J'_{k}) \,.
\end{equation}
In abstract terms, the flatness condition of the original Lax pair \(\hat{\Lax}\) can be understood as a linear combination of the \ac{eom} of the initial model,
\begin{equation}
  \dd \hat{\Lax} + \hat{\Lax} \wedge \hat{\Lax} = \sum_i \ac{eom}_i(\dd X, Y) = 0,
\end{equation}
where again $X$ are the coordinates of the torus and $Y$ the coordinates of the base. The new Lax pair ${\Lax}'$ satisfies by construction 
\begin{equation}
	\dd {\Lax}' + {\Lax}' \wedge {\Lax}' = \sum_i \ac{eom}_i(\mathcal{D}(\dd X), Y),
\end{equation}
where $\mathcal{D}$ is the $O(d,d)$ map~\eqref{eq:duality-map}. This map implements the $O(d,d)$ transformation at the level of the \ac{eom}: the set of \ac{eom} of the deformed system are equivalent to those of the initial system,
\begin{equation}
  \{\ac{eom}(\mathcal{D}(\dd X), Y)\} = \{ \ac{eom}'(\dd X', Y)\}.
\end{equation}
To see that, observe that the \ac{eom} of the sigma model are equivalent to the \ac{eom} of the doubled sigma model under the self-duality condition~\eqref{eq:self-duality}. In these terms, the $O(d,d)$ map is linear and the self-duality condition transforms covariantly under $O(d,d)$. Due to the linearity of the $O(d,d)$ map and the fact that the doubled sigma model is invariant under the map, the \ac{eom} of the transformed model are a linear combination of the initial \ac{eom} in terms of the new variables $\mathcal{D}(\dd X)$.
For example, the \ac{eom} for the adapted coordinates $X'^{i}$ (\emph{i.e.} the conservation laws for $J'_{i}$), are related to those for \(X^i\) via
\begin{equation}
0 = \dd^{2} \wt{X}'_{i} = \dd \star J'_{i} = \left(\gamma^{t}\right)_{ij} \dd^{2} X^{j} + \left(\alpha^{t}\right)_{i}{}^{k} \dd \star J_{k} = \alpha^{k}{}_{i} \dd \star J_{k}
\end{equation}
where we used~\eqref{eq:Xp-and-X-and-J}.\\
The flatness condition of the transformed Lax pair can finally be written as a linear combination of the \ac{eom} of the deformed model and is hence fulfilled on shell
\begin{equation}
  \dd {\Lax}' + {\Lax}' \wedge {\Lax}' =\sum_i \ac{eom}_i(\mathcal{D}(\dd X), Y) = \sum_i \tensor{\Lambda}{^j_i} \ac{eom}'_j (\dd X', Y)=0.
\end{equation}

This argument shows that for each Lax pair \(\Lax\) in the initial model there is a corresponding flat Lax pair \(\Lax'\) in the model resulting from the $O(d,d)$ transformation.
This is true both for symmeries in \(O(d,d; \setZ)\) and for solution-generating transformations in \(O(d,d; \setR)\).
In other words, we see that classical integrability is preserved  on general grounds under $O(d,d)$ transformations.
In the following we will present some explicit examples of integrable $O(d,d)$ transformed systems which are of general interest. 

\section{Example 1: $S^2$ and $O(1,1)$}%
\label{sec:O11}

We start with the simplest model in order to explicitly illustrate the concepts introduced in the last section, reproducing the material in~\cite{Ricci:2007eq} from a different point of view.

\paragraph{Set-up.}
The sigma model action on the two-sphere reads
\begin{equation}\begin{aligned}
\label{eq:S2-sigma}
S[\Phi, \Theta] = \frac{1}{2}\int_{\Sigma_{2}}\left[\dd \Theta \wedge \star \dd \Theta + \sin[2](\Theta)\, \dd \Phi \wedge \star\, \dd \Phi \right],
\end{aligned}\end{equation} 
where $\Phi, \Theta$ are the angle variables parametrizing the sphere. The \(SO(3)\) symmetry of the sphere has three Killing vectors:
\begin{equation}\begin{aligned}
    k_{1} &= \sin(\Phi) \del
    _{\Theta} + \cos(\Phi) \cot(\Theta) \del_{\Phi}\,,\\ 
k_{2} &= \cos(\Phi)\del_{\Theta} - \sin(\Phi) \cot(\Theta) \del_{\Phi}\,,\\
k_{3} &= \del_{\Phi}\,, 
\end{aligned}\end{equation}
satisfying
\begin{align}
[k_{A}, k_{B}] &= f_{AB}{}^{C} k_{C}, & f_{12}{}^{3} &=  1.
\end{align}
The corresponding Noether currents are given by
\begin{equation}
  \label{eq:S2-currents}
  \begin{aligned}
    J_{1} &= \sin(\Phi) \dd{\Theta} + \sin(\Theta) \cos(\Theta)\cos(\Phi)\dd{\Phi} ,\\
    J_{2} &=\cos(\Phi) \dd{\Theta} - \sin(\Theta) \cos(\Theta) \sin(\Phi) \dd{\Phi} ,\\
    J_{3} &=\sin[2](\Theta)\, \dd\Phi\,.
  \end{aligned}
\end{equation}
They are conserved, $\dd\star J_{i} =0,\ i=1,2,3$, and satisfy the flatness condition
\begin{equation}\begin{aligned}
\dd J_{i} + f_{i}{}^{jk} J_{j} \wedge J_{k} = 0\,,\qquad f_{1}{}^{23} = +1.
\end{aligned}\end{equation}
As explained above, the $J_{i}$ can be used to construct the Lax pairs $\Lax_{\lambda}$, see~\eqref{eq:LaxConnection}.
They satisfy the flatness condition 
\begin{equation}\begin{aligned}
\dd \Lax_{i} + f_{i}{}^{km} \Lax_{k} \wedge \Lax_{m} = 0\,,\qquad i,k,m=1,2,3\,.
\end{aligned}\end{equation}
Taking the path-ordered exponential of the flat currents, we can now compute infinitely many conserved non-local charges.

\paragraph{Transformations and currents.}

Now we apply an \(O(1,1)\) transformation to the sigma model above.
We will see that, while in general the \(O(3)\) symmetry is broken, we can still find a set of three conserved flat currents that imply the integrability of the deformed model.

It is natural to pick the Killing vector \(\del_\Phi\) to define the doubled torus and introduce the coordinate
\begin{equation}
  \mathbb{X} ^{I} = \pmqty{\Phi \\ \wt{\Phi}}
\end{equation}
and the corresponding generalized metric
\begin{equation}
  \mathcal{H}_{JK} = \pmqty{
    \sin[2](\Theta) & 0 \\
    0 & \frac{1}{\sin[2](\Theta)} } .
\end{equation} 
Under the action of \(g \in O(1,1)\) they transform as
\begin{align}
  \mathcal{H} &\rightarrow \mathcal{H}' =g^{t} \mathcal{H} g,\\
  \mathbb{X} & \rightarrow \mathbb{X}^{\prime }= g^{-1}\mathbb{X}.
\end{align}

It is convenient to consider the two connected components of $O(1,1; \setR)$ separately, which can be parametrized as
\begin{align}
  G_0 &= \ev{ g_0(t) = \pmqty{e^t & 0 \\ 0 & e^{-t}}} ,\\
  G_T &= \ev{ g_T(t) = \pmqty{0 & e^t  \\  e^{-t} & 0}} .
\end{align}
The first connected component includes the identity, so we expect it to describe a continuous deformation of the initial sigma model.
We will see that \(G_T\) describes the T-dual model and its deformations.

In this simple situation, the transformations in the connected component \(G_0\) are rescalings of the initial model. In fact,
\begin{align}
  \mathbb{X}^{\prime I} &= 
  \pmqty{
    e^{-t} \Phi\\
    e^{t} \wt{\Phi}
  }= \pmqty{ \Phi' \\ \wt{\Phi'}} 
  , &
  \mathcal{H}' &=
  \pmqty{
    e^{2t} \sin[2](\Theta) & 0 \\
    0 & \frac{e^{-2t}}{\sin[2](\Theta)}
  },
\end{align}
and the deformed action (in the usual polarization where the first component is physical) reads
\begin{equation}
  S_t^{0}[\Theta, \Phi']= \frac{1}{2}\int_{\Sigma}\bqty{\dd\Theta \wedge \star \dd \Theta + e^{2t} \sin[2](\Theta) \dd \Phi' \wedge \star \dd \Phi'},
\end{equation}
which locally still describes a two-sphere.

The situation is more interesting for \(G_T\).
In this case,
\begin{align}
  \label{eq:T-dual-doubled-S2}
  \mathbb{X}^{\prime I} &= 
  \pmqty{
     e^{t} \wt{\Phi}\\
     e^{-t} \Phi
  } = \pmqty{ \Phi' \\ \wt{\Phi'}}, & 
  \mathcal{H}' &= 
  \pmqty{
     \frac{e^{-2t}}{\sin[2](\Theta)} & 0 \\
    0 & e^{2t}\sin[2](\Theta)
  },
\end{align}
and the deformed sigma model reads
\begin{equation}
  \label{eq:dual-sigma-model-S2}
  S'_{t}[\Theta, \Phi'] = \frac{1}{2}\int_{\Sigma}\bqty{\dd \Theta \wedge \star \dd \Theta + \frac{e^{-2t}}{\sin[2](\Theta)} \dd \Phi^{\prime }\wedge \star \dd \Phi'}.
\end{equation}
It is easy to recognize this as a local rescaling of the T-dual model.

The important observation is that this model has only one isometry, corresponding to the Killing vector \(\del_{\Phi'}\), so Noether's construction would only lead to one conserved current.
On the other hand we know that this system is related to the original \(S^2\) sigma model by an \(O(1,1)\) transformation.
As we have seen in Section~\ref{sec:Odd-currents}, we can simply follow the action of this transformation on the three conserved currents in Eq.~(\ref{eq:S2-currents}). 
To do so, we need to
\begin{itemize}
\item Find an \(SO(3)\) transformation of the initial Lax pair to find a gauge in which it is manifestly invariant under \(\del_{\Phi}\), \emph{i.e.} does not depend on \(\Phi\) but only on \(\dd \Phi\). We use
  \begin{equation}
    \hat{\Lax} = h^{-1} \Lax h + h^{-1} \dd{h}
  \end{equation}
  with 
  \begin{equation}
  	h = \pmqty{\cos \Phi & \sin \Phi & 0\\
  	-\sin \Phi & \cos \Phi & 0 \\
  	0 & 0 & 1 }
  \end{equation}
  and
  \begin{equation}
    \begin{aligned}
      \hat{\Lax}^{1} &= \sin(\Theta)\cos(\Theta)(a\dd\Phi + b \star \dd\Phi)\,,\\[2pt]
      \hat{\Lax}^{2} &= a \dd \Theta + b \star \dd \Theta\,,\\[2pt]
      \hat{\Lax}^{3} &=\sin[2](\Theta)(a \dd \wt{\Phi} + b \star \dd \wt{\Phi}) - 2\dd \Phi\,,
    \end{aligned}
  \end{equation}
  where $a$ and $b$ were given in Eq.~\eqref{eq:LaxFlat}.
\item Impose the self-duality condition
  \begin{equation}
    \dd{\mathbb{X}^{I}} =  L^{IJ} \mathcal{H}_{JK} \star \dd{\mathbb{X}^{K}}\,,
  \end{equation}
 which in our case reads
 \begin{equation}
   \label{eq:self-dual}
   \pmqty{
     \dd \Phi\\
     \dd \wt{\Phi}}
   = \pmqty{
       \frac{1}{\sin[2](\Theta)} \star \dd \wt{\Phi}\\
       \sin[2](\Theta) \star \dd \Phi
     }.
 \end{equation}
\end{itemize}
Then we find for the dual Lax pair \(\Lax'^{1}(\dd \Phi') = \hat{\Lax}^{1}(\dd \Phi \rightarrow \dd \Phi')\):
\begin{multline}
  \hat{\Lax}^{1} = \sin(\Theta) \cos(\Theta)(a \dd{\Phi} + b \star \dd{\Phi}) = \sin(\Theta) \cos(\Theta) \pqty{a \frac{1}{\sin[2](\Theta)} \star \dd{\wt{\Phi}} + b \frac{1}{\sin[2](\Theta)} \dd{\wt{\Phi}}} \\
  = e^{-t} \frac{\cos(\Theta)}{\sin(\Theta)} \pqty{ a * \dd{\Phi'} + b \dd{\Phi'}}= \Lax'^{1}(\dd\Phi'),  
\end{multline}
where we have first imposed the self-duality condition and then used Eq.~(\ref{eq:T-dual-doubled-S2}) to relate \(\wt{\Phi}\) to the redefined coordinate \(\Phi'\) of the deformed model.

Repeating the construction for the other currents we find that the sigma model in Eq.~(\ref{eq:dual-sigma-model-S2}) admits the following three conserved currents:
\begin{equation}
  \begin{aligned}
    \Lax'^{1} &= e^{-t} \frac{\cos(\Theta)}{\sin(\Theta)} \pqty{ a \star \dd{\Phi'} + b \dd{\Phi'}} ,\\
    \Lax'^{2} &= a \dd{\Theta} + b \star \dd{\Theta} ,\\
    \Lax'^{3} &= e^{-t} \pqty{a \star \dd{\Phi'} + b \dd{\Phi'}- \frac{2}{\sin[2](\Theta)} \star \dd{\Phi'}} .
  \end{aligned}
\end{equation}
These currents are flat on shell as expected.
They are precisely the non-local T-dual currents discussed in~\cite{Ricci:2007eq}. As we have seen, the generalization provided by considering the full \(O(1,1)\) group is limited to local rescalings.

The strength of our general formalism is that it can be applied to larger \(O(d,d)\) groups, which in general describe non-trivial deformations beyond T-duality, as we will see in the next section.

\section{Example 2: $S^{3}$ and $O(2,2)$}%
\label{sec:O22}

In this section, we study the group of $O(2,2)$ transformations for the sigma model on the three-sphere with non-zero $H$-flux. 

\subsection{WZNW model on the three-sphere}
Let us start with the \ac{wznw} model on the group \(SU(2)\):
\begin{equation}
  \begin{aligned}
    S[g] &= -\frac{1}{4}\int_{\Sigma} \Tr[ g^{-1} \dd g \wedge \star g^{-1} \dd g ] + \frac{i\kappa}{3!} \int_{\mathcal{V}}\Tr[ g^{-1} \dd g \wedge g^{-1} \dd g \wedge g^{-1} \dd g ]\\ 
    &= -\frac{1}{4}\int_{\Sigma} \Tr[ \omega_{\LL} \wedge \star \omega_{\LL} ]+ \frac{i \kappa}{3!} \int_{\mathcal{V}} \Tr [ \omega_{\LL} \wedge \omega_{\LL} \wedge \omega_{\LL} ]\\
    &= -\frac{1}{4}\int_{\Sigma} \Tr [ \omega_{\RR}\wedge \star \omega_{\RR} ] - \frac{i \kappa}{3!} \int_{\mathcal{V}} \Tr [ \omega_{\RR} \wedge \omega_{\RR} \wedge \omega_{\RR} ],
  \end{aligned}
\end{equation}
where \(\kappa=1\) at the conformal point and the left/right-invariant Maurer--Cartan one-forms $j_{\LL/\RR}$ are defined as
\begin{align}
\label{eq:MC-form}
\omega_{\LL} &= g^{-1} \dd g, &  \omega_{\RR} &= - \dd g\,g^{-1}
\end{align}
for \(g \in SU(2)\).
By construction, the currents satisfy the Mauer--Cartan equations,
\begin{equation}
\dd \omega + \omega \wedge \omega = 0\,.
\end{equation}
The variation of the action is given by
\begin{equation}\begin{aligned}
\delta S  &= +\frac{1}{2}\int_{\Sigma} \mathrm{Tr}\,(g^{-1} \delta g)\left(\dd \star \omega_{\LL} - i \kappa  \dd{ \omega_{\LL}} \right)\\
&=-\frac{1}{2}\int_{\Sigma}\mathrm{Tr}\,(\delta g\, g^{-1}) \left(  \dd \star \omega_{\RR} + i \kappa \dd{ \omega_{\RR}} \right)\,,
\end{aligned}\end{equation}
which leads to the \ac{eom}:
\begin{align}
\dd \star \omega_{\LL} - i \kappa \dd{ \omega_{\LL}} &= 0, & \dd \star \omega_{\RR} + i \kappa  \dd{ \omega_{\RR}} &= 0.
\end{align}
The Noether currents associated to the \(SU(2) \times SU(2)\) global symmetry are given by %
\begin{align}
\label{eq:WZNW-current}
J_{\LL} &= \omega_{\LL} - i \kappa \star \omega_{\LL}, & J_{\RR} &= \omega_{\RR} + i \kappa \star \omega_{\RR}.
\end{align}
We see that they are not only conserved but also flat for any value of \(\kappa\):
\begin{equation}\begin{aligned}
\dd J_{\LL} + J_{\LL} \wedge J_{\LL} &= (\dd \omega_{\LL} - i\kappa \dd{ \star \omega_{\LL}}) + (1 + \kappa^2) \omega_{\LL} \wedge \omega_{\LL}\\
&= (1 + \kappa^2 ) (\dd \omega_{\LL} + \omega_{\LL} \wedge \omega_{\LL}) = 0 \,.
\end{aligned}\end{equation}
The same can be verified for \(j_{\RR}\).
From the conserved and flat currents we can construct the Lax pairs:
\begin{equation}
  \begin{aligned}
    \Lax_{\LL} &= a\, J_{\LL} +b \star J_{\LL} = - \left( (i \kappa b-a) + (i \kappa a-b) \star \right) \omega_{\LL},\\
    \Lax_{\RR} &= a\, J_{\RR} +b \star J_{\RR} = +\left( (i \kappa b+a) + (i \kappa a + b) \star \right) \omega_{\RR},
  \end{aligned}
\end{equation}
where $a, b$ contain the spectral parameter $\lambda$ as before, and are given by
\begin{align}
a &= \frac{1}{2}(1 \pm \cosh(\lambda) ), & b &= \frac{1}{2} \sinh(\lambda),
\end{align}
and satisfy $a^{2} - b^{2} - a = 0$.
Using the Lax pairs, the infinite number of conserved charges can be constructed in the usual way (see Sec.~\ref{sec:Odd-currents}).

\bigskip

From now on we will set $\kappa = 1$ and look at deformations of the conformal model for ease of notation.  It is convenient to pick an explicit parametrization for $g \in SU(2)$:
\begin{equation}
  \label{eq:SU2-element}
  g = e^{- (\zeta_{1}+\zeta_{2}) T_{2}} e^{\eta\, T_{1}} e^{+(\zeta_{1} - \zeta_{2}) T_{2}}, 
\end{equation}
where the generators $T_{\alpha}$, $\alpha=1,2,3$ are defined in terms of the usual Pauli matrices
\begin{align}
  T_{\alpha} &= -\frac{i}{2}\sigma_{\alpha}, & \alpha&=1,2,3,
\end{align}
and satisfy
\begin{align}
  \comm{T_{\alpha}}{T_{\beta}} &= \epsilon_{\alpha\beta\gamma} T_{\gamma}, & \Tr(T_{\alpha}T_{\beta}) &= -\frac{1}{2}\delta_{\alpha\beta},
\end{align}
where $\epsilon_{123} = 1$.

The metric and $H$-flux are read off from
\begin{equation}
  \begin{aligned}
    -\frac{1}{4}\mathrm{Tr}\left[ g^{-1}\dd g \wedge \star g^{-1} \dd g\right] &=\frac{1}{8} \dd \eta \wedge \star \dd \eta + \frac{1}{2} \sin[2](\frac{\eta}{2}) \dd \zeta_{1} \wedge \star \dd \zeta_{1} \\
    &\hspace{3cm} + \frac{1}{2} \cos[2](\frac{\eta}{2} ) \dd \zeta_{2} \wedge \star \dd \zeta_{2},\\
    \frac{1}{3!}\mathrm{Tr}\left[g^{-1} \dd g \wedge g^{-1} \dd g \wedge g^{-1} \dd g \right] &= - \sin(\frac{\eta}{2})\cos(\frac{\eta}{2}) \dd \zeta_{1} \wedge \dd \zeta_{2} \wedge \dd \eta,
  \end{aligned}
\end{equation}
which, for example, leads to the equation of motion for $\eta$ 
\begin{equation}
\label{eq:EOM-eta}
\dd \star \dd \eta - \sin(\eta) (\dd \zeta_{1} \wedge \star \dd \zeta_{1} - \dd \zeta_{2} \wedge \star \dd \zeta_{2} - 2 i \dd \zeta_{1} \wedge \dd \zeta_{2}) = 0\,.
\end{equation}
It is also convenient to decompose the Maurer-Cartan forms on the basis of the $T_{\alpha}$ in the following. Using the $SU(2)$ element~\eqref{eq:SU2-element}, we explicitly write down~\eqref{eq:MC-form} as
\begin{equation}\begin{aligned}
\label{eq:left-invariant}
\omega_{\LL}^{1} &=\sin(\zeta_{-})\sin(\eta) \dd \zeta_{+} - \cos(\zeta_{-}) \dd \eta\,,\\
\omega_{\LL}^{2} &=\dd \zeta_{-} - \cos(\eta) \dd \zeta_{+} = 2 \left(\sin[2](\frac{\eta}{2})\dd \zeta_{1} - \cos[2](\frac{\eta}{2})\dd \zeta_{2}\right)\,,\\
\omega_{\LL}^{3} &= -\cos(\zeta_{-}) \sin(\eta) \dd \zeta_{+} - \sin(\zeta_{-}) \dd \eta\\ 
\end{aligned}\end{equation}
while
\begin{equation}\begin{aligned}
\label{eq:right-invariant}
\omega_{\RR}^{1} &=-\sin(\zeta_{+})\sin(\eta) \dd \zeta_{-} + \cos(\zeta_{+}) \dd \eta\,,\\
\omega_{\RR}^{2} &=\dd \zeta_{+} - \cos(\eta) \dd \zeta_{-} = 2 \left( \sin[2](\frac{\eta}{2}) \dd \zeta_{1} + \cos[2](\frac{\eta}{2})\dd \zeta_{1} \right)\,,\\
\omega_{\RR}^{3} &= \cos(\zeta_{+}) \sin(\eta) \dd \zeta_{-} + \sin(\zeta_{+}) \dd \eta\,,
\end{aligned}\end{equation}
where we introduced the coordinates
\begin{equation}
  \zeta_{\pm} = \zeta_{1} \pm \zeta_{2} .
\end{equation}
In terms of this decomposition, the flatness condition reads
\begin{equation}
  \dd \omega^{\alpha} + \frac{1}{2} \epsilon^{\alpha}{}_{\beta\gamma}\, \omega^{\beta} \wedge \omega^{\gamma} = 0\,,\qquad\alpha, \beta, \gamma= 1,2,3\,.
\end{equation}
The corresponding Killing vectors $k^{\alpha}_{\LL/\RR}\,,\ \alpha=1,2,3$ to the above one-forms are computed as
\begin{equation}
\begin{aligned}
k_{\LL}^{1} &= +\csc(\eta)\sin(\zeta_{-})\partial_{\zeta_{+}} + \cot(\eta)\sin(\zeta_{-})\partial_{\zeta_{-}} - \cos(\zeta_{-})\partial_{\eta},\\
k_{\LL}^{2} &=  \partial_{\zeta_{-}},\\
k_{\LL}^{3} &=  - \csc(\eta)\cos(\zeta_{-}) \partial_{\zeta_{+}} - \cot(\eta)\cos(\zeta_{-})\partial_{\zeta_{-}}- \sin(\zeta_{-})\partial_{\eta}
\end{aligned}
\end{equation} 
whereas 
\begin{equation}
\begin{aligned}
k_{\RR}^{1} &= -\cot(\eta)\sin(\zeta_{+})\partial_{\zeta_{+}} - \csc(\eta)\sin(\zeta_{+})\partial_{\zeta_{-}} + \cos(\zeta_{+})\partial_{\eta},\\
k_{\RR}^{2} &=  \partial_{\zeta_{+}},\\
k_{\RR}^{3} &=  +\cot(\eta)\cos(\zeta_{+}) \partial_{\zeta_{+}} + \csc(\eta)\cos(\zeta_{+}) \partial_{\zeta_{-}} + \sin(\zeta_{+}) \partial_{\eta}.
\end{aligned}
\end{equation} 
As is clear above, $k_{\LL}^{2}$ and $k_{\RR}^{2}$ are commuting Killing vectors in our choice of coordinates. The associated conserved currents are 
\begin{align}
   J_{\LL}^{2} &= 2  (J_{1} - J_{2})\,, &  J_{\RR}^{2} &= 2  (J_{1} + J_{2})\,,
\end{align}
where $J_{1}$ and $J_{2}$ are $U(1) \times U(1)$ Noether currents for the Killing vectors $\partial_{\zeta_{1}}$ and $\partial_{\zeta_{2}}$\,, respectively:
\begin{equation}
\label{eq:U(1)-currents}
\begin{aligned}
J_{1} &= \sin[2](\frac{\eta}{2}) \dd \zeta_{1} - i \sin[2](\frac{\eta}{2}) \star \dd \zeta_{2}\,,\\
J_{2} &= \cos[2](\frac{\eta}{2}) \dd \zeta_{2} + i \sin[2](\frac{\eta}{2}) \star \dd \zeta_{1}\,. 
\end{aligned}
\end{equation}
In the following we will pick $\zeta_{1}$ and $\zeta_{2}$ to be adapted coordinates in the doubled formalism, so that we can study the corresponding action of \(O(2,2)\)\,.

\subsection{Doubled formalism}

We start by introducing the doubled coordinates
\begin{equation}
  \mathbb{X}^{I} = \pmqty{ \zeta_{1} \\ \zeta_{2} \\  \wt{\zeta}_{1} \\ \wt{\zeta}_{2}}
\end{equation}
and read from the sigma model the generalized metric
\begin{equation}
  \mathcal{H} = \pmqty{G- BG^{-1}B & BG^{-1} \\ - G^{-1} B & G^{-1}}
  = 
  \pmqty{
1 & 0 & 0 & 1 \\
0 & +\cot^{2}\left(\frac{\eta}{2}\right) & - \cot^{2}\left(\frac{\eta}{2}\right) & 0 \\[10pt]
0 & - \cot^{2}\left(\frac{\eta}{2}\right) & + \csc^{2}\left(\frac{\eta}{2}\right) & 0 \\
1 & 0 & 0 & \sec^{2}\left(\frac{\eta}{2}\right)
}.
\end{equation}
The two ingredients of our construction are the choice of an appropriate gauge for the flat currents and the \(O(2,2)\)  map.
The gauge choice depends only on the torus that we have picked and remains the same for all the transformations in \(O(2,2)\).

\paragraph{Gauge choice.}
For $\Lax_{\LL}$, we pick
\begin{equation}
h = e^{-\zeta_{-} T_{2}}\,,
\end{equation}
so that the gauged Lax pairs decompose as
\begin{equation}\begin{aligned}
\hat{\Lax}^{1}_{\LL} &=  + \left( (i b - a )  + (i a - b) \star \right)\dd \eta\,,\\
\hat{\Lax}^{2}_{\LL} &=  -\left( (i b - a ) + ( i  a - b ) \star \right) ( \dd \zeta_{-} - \cos(\eta) \dd \zeta_{+})- \dd \zeta_{-}  \,,\\
\hat{\Lax}^{3}_{\LL} &=  +\left( (i b - a) + (i a - b) \star \right) \sin (\eta) \dd \zeta_{+}\,.
\end{aligned}\end{equation}
On the other hand, for ${J}_{\RR}$, we choose
\begin{equation}
  h = e^{- \zeta_{+} T_{2}}\,,
\end{equation}
so that
\begin{equation}
  \begin{aligned}
    \hat{\Lax}^{1}_{\RR} &=  +\left( (i b + a ) + (i a + b) \star \right) \dd \eta\,,\\
    \hat{\Lax}^{2}_{\RR} &=  + \left( (i b + a ) + (i a + b) \star \right)  (\dd \zeta_{+} - \cos(\eta) \dd \zeta_{-})- \dd \zeta_{+} \,,\\
    \hat{\Lax}^{3}_{\RR} &=  +\left( (i b + a) + (i a + b) \star \right)\sin(\eta) \dd \zeta_{-}\,.
  \end{aligned}
\end{equation}
The first components $\hat{\Lax}^1_{\LL/\RR}$ do not have components in the directions \(\dd{\zeta_\pm}\),  and they will not be affected by $O(2,2)$ transformations.

Finally let us observe explicitly the connections of the gauged Lax pairs. In the following some example of special interest will be discussed. Using the $U(1)^{2} $ conserved currents~\eqref{eq:U(1)-currents}\,, we compute
\begin{equation}
\begin{aligned}
\dd \hat{\Lax}^{1}_{\LL} + \hat{\Lax}^{2}_{\LL} \wedge \hat{\Lax}^{3}_{\LL}  &=(ia-b) \left[ \dd \star \dd \eta - \sin(\eta) \left(\dd \zeta_{1} \wedge \star \dd \zeta_{1} - \dd \zeta_{2} \wedge \star \dd \zeta_{2} - 2i \dd \zeta_{1} \wedge \dd \zeta_{2}\right)\right] ,\\
\dd \hat{\Lax}^{2}_{\LL} + \hat{\Lax}^{3}_{\LL} \wedge \hat{\Lax}^{1}_{\LL} &=- 2 (ia-b) \left( \dd \star J_{1} - \dd \star J_{2} \right)\,,\\
\dd \hat{\Lax}^{3}_{\LL} + \hat{\Lax}^{1}_{\LL} \wedge \hat{\Lax}^{2}_{\LL} &= 2(ia-b)\left[\cot(\frac{\eta}{2})\dd \star J_{1} + \tan(\frac{\eta}{2}) \dd \star J_{2}  \right]
\end{aligned}
\end{equation}
as well as
\begin{equation}
\begin{aligned}
\dd \hat{\Lax}^{1}_{\RR} + \hat{\Lax}^{2}_{\RR} \wedge \hat{\Lax}^{3}_{\RR}  &=(ia+b) \left[ \dd \star \dd \eta - \sin(\eta) \left(\dd \zeta_{1} \wedge \star \dd \zeta_{1} - \dd \zeta_{2} \wedge \star \dd \zeta_{2} - 2i \dd \zeta_{1} \wedge \dd \zeta_{2}\right)\right] ,\\
\dd \hat{\Lax}^{2}_{\RR} + \hat{\Lax}^{3}_{\RR} \wedge \hat{\Lax}^{1}_{\RR} &=  2 (ia+b) \left( \dd \star J_{1} + \dd \star J_{2} \right)\,,\\
\dd \hat{\Lax}^{3}_{\RR} + \hat{\Lax}^{1}_{\RR} \wedge \hat{\Lax}^{2}_{\RR} &= +2(ia+b)\left[\cot(\frac{\eta}{2})\dd \star J_{1} - \tan(\frac{\eta}{2}) \dd \star J_{2}  \right]\,,
\end{aligned}
\end{equation}
which obviously vanish under the \ac{eom}~\eqref{eq:EOM-eta} and the conservation of $J_{1}$ and $J_{2}$.

\subsection{Marginal deformations}
Current--current deformations of \ac{wznw} models are described by transformations \(\mathcal{O} \in O(d) \times O(d) \subset O(d,d)\).
In the case of a \ac{wznw} model on a compact Lie group, all maximal Abelian subgroups are pairwise conjugated by inner automorphisms so the complete deformation space is \(D = O(r,r)/(O(r)\times O(r))\), where \(r\) is the rank of the group (see \emph{e.g.}~\cite{Forste:2003km}).
In our case, there is only one possible deformation of this kind, which we will call \(J \bar J \)-deformation.

To realize this deformation, we can consider for example the element \(g(\alpha) \in O(2,2)\) written as
\begin{equation}
\label{eq:rotation-element}
  g(\alpha) = \frac{1}{2}\pmqty{
    R(\alpha)+S(\alpha) & R(\alpha)-S(\alpha) \\
      R(\alpha)-S(\alpha) & R(\alpha)+S(\alpha)} , 
\end{equation}
where 
\begin{equation}
  R(\alpha) = S(\alpha)^{-1} = \pmqty{
    \cos(\alpha) & \sin(\alpha) \\
    - \sin(\alpha) & \cos(\alpha)} .
\end{equation}
The $J\bar{J}$-deformation results from performing~\cite{Giveon:1993ph,Hassan:1992gi}
\begin{enumerate}
\item an $O(2)\times O(2)$ rotation $g(\alpha)$ given by~\eqref{eq:rotation-element},
\item a diffeomorphism $g_{A}$~\eqref{eq:diffeo} characterized by
\begin{equation}
A(\alpha) = \pmqty{
	\cos(\alpha) + \sin(\alpha)  & 0 \\
	0 & \cos(\alpha)
	}\,,
\end{equation}
\item a $B$-shift $g_{B}$~\eqref{eq:B-shift} given by 
\begin{equation}
\dd \Lambda(\alpha) = 
	\pmqty{
	0 & \cos(\alpha)(\sin(\alpha)-\cos(\alpha)) \\
	- \cos(\alpha)(\sin(\alpha)-\cos(\alpha))  & 0
	}\,.
\end{equation} 
\end{enumerate}
In summary, the $O(2,2)$ transformation corresponding to the $J\bar J$-deformation is given by
\begin{equation}
\label{eq:JJbar-element}
g_{J \bar{J}}(\alpha) =  g(\alpha) g_{\mathrm{diff}}(A(\alpha)) g_{B}(\dd \Lambda(\alpha))
= \pmqty{
	1 & 0 & 0 & \tan(\alpha) \\
	0 & \frac{\cos(\alpha)}{\cos(\alpha)+\sin(\alpha)}& - \frac{\sin(\alpha)}{\cos(\alpha)+\sin(\alpha)} & 0 \\
	0 & \frac{\cos(\alpha)}{\cos(\alpha)+\sin(\alpha)} & \frac{\cos(\alpha)}{\cos(\alpha)+\sin(\alpha)} & 0 \\
	-1 & 0 & 0 & 1 
	}.
\end{equation}
Then the generalized metric transforms as
\begin{equation}
\begin{aligned}
  \mathcal{H}^{\prime} &= g_{J \bar{J}}(\alpha)^{t} \mathcal{H}g_{J \bar{J}}(\alpha)\\
  &=
  \pmqty{
   \tan[2]( \frac{\eta}{2} ) & 0 & 0 & - \tan[2]( \frac{\eta}{2} ) \\
   0 & \frac{\cos[2](\alpha)}{(\cos(\alpha) + \sin(\alpha))^{2}} & \frac{\cos[2](\alpha)}{(\cos(\alpha) + \sin(\alpha))^{2}} & 0\\
   0 & \frac{\cos[2](\alpha)}{(\cos(\alpha) + \sin(\alpha))^{2}} & \frac{\Delta_{1}}{(\cos(\alpha) + \sin(\alpha))^{2}\sin[2]( \frac{\eta}{2} ) } & 0 \\
   -\tan[2]( \frac{\eta}{2} )  & 0 & 0 & \frac{\Delta_{1}}{\cos[2](\alpha)\cos[2]( \frac{\eta}{2})}
   		}\,,
\end{aligned}
\end{equation}
where 
\begin{equation}
\Delta_{1} = \cos[2](\alpha) + \cos[2](\frac{\eta}{2})\sin(\alpha)( \sin(\alpha) + 2\cos(\alpha))\,.
\end{equation}
This leads to the following deformed sigma-model:
\begin{multline}
S'_{\alpha}[\eta, \zeta'_{1}, \zeta'_{2}] = \frac{1}{2}\int \frac{1}{4} \dd \eta \wedge \star \dd \eta + \frac{1}{\Delta_{1}}(\cos(\alpha) + \sin(\alpha))^{2} \sin[2](\frac{\eta}{2})\dd \zeta'_{1} \wedge \star \dd \zeta'_{1}\\
 + \frac{1}{\Delta_{1}}\cos[2](\alpha)\cos[2]( \frac{\eta}{2}) \dd \zeta'_{2} \wedge \star \dd \zeta'_{2} - \frac{2i}{\Delta_{1}} \cos[2](\alpha)\sin[2]( \frac{\eta}{2} ) \dd \zeta'_{1} \wedge \dd \zeta'_{2}\,.
\end{multline}
The \ac{eom} for $\zeta'_{1,}\,, \zeta'_{2}$ can be expressed as
\begin{equation}
\dd \star J_{1}(\alpha) = 0 \,,\qquad \dd \star J_{2}(\alpha) = 0\,,
\end{equation}
where we define the conserved currents of the dual model to be
\begin{align}
J_{1}(\alpha) &= \frac{1}{\Delta_{1}}\left[ (\cos(\alpha) + \sin(\alpha))^{2} \sin[2](\frac{\eta}{2}) \dd \zeta'_{1} - i \cos[2](\alpha) \sin[2](\frac{\eta}{2}) \star \dd \zeta'_{2}\right]\,,\\
J_{2}(\alpha) &= \frac{1}{\Delta_{1}}\cos[2](\alpha)\left[\cos[2](\frac{\eta}{2})\dd \zeta'_{2} + i \sin[2](\frac{\eta}{2}) \star \dd \zeta'_{1}\right]\,.
\end{align}
They obviously correspond to the $U(1) \times U(1)$ isometries along the Killing vectors $\partial_{\zeta'_{1}}$ and $\partial_{\zeta'_{1}}$\,, respectively. Note that 
\begin{equation}
S'_{\alpha}[\eta, \zeta'_{1}, \zeta'_{2}] = \frac{1}{2} \int \frac{1}{4} \dd \eta \wedge \star \dd \eta + \dd \zeta'_{1} \wedge \star J_{1}(\alpha) + \dd \zeta'_{2} \wedge \star J_{2}(\alpha) .
\end{equation}
At order $O(\alpha)$, the action is approximated as
\begin{equation}
\begin{aligned}
 S_{0}[\eta, \zeta'_{1}, \zeta'_{2}] + \frac{\alpha}{2}\int 2 \sin[4](\frac{\eta}{2}) \dd \zeta'_{1} \wedge \star\dd \zeta'_{1} - 2 \cos[4](\frac{\eta}{2}) \dd \zeta'_{2} \wedge \star \dd \zeta'_{2}+ i \sin[2](\eta) \dd \zeta'_{1} \wedge \dd \zeta'_{2}\\
= S_{0}[\eta, \zeta'_{1}, \zeta'_{2}] - \frac{i\alpha}{2}\int \frac{1}{4} J^{2}_{\LL}(\zeta') \wedge (1+ i\star ) J^{2}_{\RR}(\zeta')\,,
\end{aligned}
\end{equation}
where $S_{0}$ is the undeformed action whereas $J^{2}_{\LL/\RR}(\zeta')$ are undeformed conserved currents~\eqref{eq:WZNW-current} with their arguments replaced by $\zeta'_{1/2}$\,.

To construct the flat currents of the $J\bar{J}$-deformed model, let us apply the inverse map of the duality automorphism~\eqref{eq:Odd-inverse}. It leads to the following relations:
\begin{equation}
\label{eq:Odd-inverse-JJbar}
\begin{aligned}
\dd \zeta_{1} ={}& \dd \zeta'_{1} + \tan(\alpha) \star J_{2}(\alpha) \,,\\
\dd \zeta_{2} ={}& \frac{1}{1 + \tan(\alpha)} \left[\dd \zeta'_{2} - \tan(\alpha) \star J_{1}(\alpha)\right]\,.
\end{aligned}
\end{equation}
We can now explicitly give the deformed Lax pair $\Lax' = \hat{\Lax}(\dd \zeta_{1/2} \rightarrow \dd \zeta'_{1/2})$:
\begin{equation}\begin{aligned}
\Lax'^{1}_{\LL} ={}& \biggl( (i   b - a ) + (i   a - b) \star \biggr) \dd \eta\,,\\
\Lax'^{2}_{\LL} ={}& -2\biggl((i  b-a) + (i  a-b) \star \biggr)
\biggl[ 
	\pmqty{ \sin[2](\frac{\eta}{2}) & \frac{-\cos[2](\frac{\eta}{2})}{1+ \tan(\alpha)} } 
	\pmqty{ \dd \zeta'_{1} + \tan(\alpha) \star J_{2}(\alpha)\\
			  \dd \zeta'_{2} - \tan(\alpha) \star J_{1}(\alpha) }
\biggr] \\
& - \dd \zeta'_{1} - \tan(\alpha) \star J_{2}(\alpha) + \frac{1}{1+\tan(\alpha)}\left( \dd \zeta'_{2} - \tan(\alpha) \star J_{1}(\alpha)\right)\,,\\
\Lax'^{3}_{\LL} ={}& \biggl((i  b-a) + (i  a-b) \star\biggr) \sin(\eta)\biggl[ \dd \zeta'_{1} + \tan(\alpha) \star J_{2}(\alpha) + \frac{\dd \zeta'_{2} - \tan(\alpha) \star J_{1}(\alpha)}{1+\tan(\alpha)} \biggr] ,
\end{aligned}\end{equation}
whereas
\begin{equation}\begin{aligned}
\Lax'^{1}_{\RR} ={}& \biggl( (i  b + a ) + (i  a + b) \star \biggr)\dd \eta\,,\\
\Lax'^{2}_{\RR} ={}& 2\biggl( (i b + a) + (i a + b)  \star \biggr)
\biggl[ 
	\pmqty{ \sin[2](\frac{\eta}{2}) & \frac{+\cos[2](\frac{\eta}{2})}{1+ \tan(\alpha)} } 
	\pmqty{ \dd \zeta'_{1} + \tan(\alpha) \star J_{2}(\alpha)\\
			  \dd \zeta'_{2} - \tan(\alpha) \star J_{1}(\alpha) }
\biggr]\\
& - \dd \zeta'_{1} - \tan(\alpha) \star J_{2}(\alpha) - \frac{1}{1+\tan(\alpha)}\left( \dd \zeta'_{2} - \tan(\alpha) \star J_{1}(\alpha)\right)\,,\\
\Lax'^{3}_{\RR} ={}& + \biggl( (i b + a) + (i a + b)  \star \biggr) \sin(\eta) \biggl[ \dd \zeta'_{1} + \tan(\alpha) \star J_{2}(\alpha) - \frac{\dd \zeta'_{2} - \tan(\alpha) \star J_{1}(\alpha)}{1+\tan(\alpha)}\biggr] ,
\end{aligned}\end{equation}
where the inner product of the two-component vectors is used in $\mathcal{L}'^{2}_{\LL/\RR}$.
Based on our general arguments in Section~\ref{sec:Odd-currents}, we were thus able to explicitly give the Lax pair of the $J\bar J$-deformed sigma model, which guarantees its classical integrability.

\subsection{TsT transformation}
TsT transformations~\cite{Alday:2005ww,Frolov:2005dj,Matsumoto:2014ubv,Matsumoto:2014nra,Matsumoto:2014gwa,Matsumoto:2015ypa,vanTongeren:2015soa,Osten:2016dvf,Delduc:2017fib} can be also interpreted as $O(d,d)$ transformations and conveniently described  using the doubled formalism.\footnote{See also a recent work~\cite{Catal-Ozer:2019tmm}.} 
In this section we will derive explicitly the six non-local flat currents of the resulting model.

Given the two-torus generated by $(\zeta_{1}, \zeta_{2})$, the corresponding TsT transformation consists of
\begin{equation}
  \text{TsT transformation}
  =
  \begin{cases}
    \text{1.\ a T-duality along $\zeta_{1}$ direction}\,,\\
    \text{2.\ a shift $\zeta_{2}$ by $\alpha\,\wt{\zeta}_{1}$,}\\
    \text{3.\ a T-duality along $\wt{\zeta}_{1}$ direction}\,.
  \end{cases} 
\end{equation}
The T-duality along $\zeta_1$ is realized by the matrix
\begin{equation}
  g_{T_{1}} =
  \pmqty
  {
    0 &0 &1 &0 \\
    0 &1 &0 &0 \\
    1 &0 &0 &0 \\
    0 &0 &0 &1
  }\,,
\end{equation}
and the doubled coordinates transform as
\begin{equation}
  (g_{T_{1}}^{-1})^{I}{}_{J} \dd \mathbb{X}^{J} = \pmqty{\dd{\wt{\zeta}_{1}} \\ \dd{\zeta_2} \\ \dd{\zeta_1} \\ \dd \wt{\zeta}_{2} }.
\end{equation}
Next, we  perform the shift transformation,
\begin{equation}
  \dd{\zeta_2} \rightarrow \dd{\zeta_2} + \alpha \dd{\zeta_1} .
\end{equation}
In order to have a consistent transformation in \(O(2,2)\), this shift must be supplemented with an opposite shift of the dual coordinates.
Then we have
\begin{equation}
  S_{\alpha} = 
  \pmqty
  {
    1 &0 &0 &0 \\
    -\alpha &1 &0 &0 \\
    0 &0 &1 &\alpha \\
    0 &0 &0 &1
  }\,. 
\end{equation}
The complete TsT transformation is realized as
\begin{equation}
  g_{\mathrm{TsT}} = g_{T_{1}} S_{\alpha} g_{T_{1}} ,
\end{equation} 
which is expressed in components as
\begin{equation}
\label{eq:TsT-element}
  g_{\mathrm{TsT}} = 
  \pmqty
  {
    1 &0 &0 &\alpha \\
    0 &1 &-\alpha &0 \\
    0 &0 &1 &0 \\
    0 &0 &0 &1
  } .
\end{equation}
It clearly lives in the component of $O(2,2)$ connected to the identity.
In the classification of Sec.~\ref{sec:Odd-group} this is a \(\beta\)-transformation with bi-vector
\begin{equation}
\beta \equiv \frac{1}{2}\beta^{ij}\partial_{X^{i}} \wedge \partial_{X^{j}} = \alpha \partial_{\zeta_{1}} \wedge \partial_{\zeta_{2}}\,,
\end{equation}
which can be identified with an Abelian classical $r$-matrix in terms of a Yang--Baxter deformation.

Using the $O(2,2)$ element~\eqref{eq:TsT-element}, we compute the deformed sigma model as\footnote{This background is related to (3.76) in \cite{Plauschinn:2018wbo} under the suitable coordinate changes. See also \cite{Plauschinn:2014nha}.
}
\begin{equation} 
\begin{aligned}
\label{eq:TsT-action}
S'_{\lambda}[\eta, \zeta'_{1}, \zeta'_{2}] &= \frac{1}{2} \int \frac{1}{4} \dd \eta \wedge \star \dd \eta + \frac{1}{\Delta_{2}}\sin[2](\frac{\eta}{2}) \dd \zeta'_{1} \wedge \star \dd  \zeta'_{1} + \frac{1}{\Delta_{2}} \cos[2](\frac{\eta}{2}) \dd \zeta'_{2} \wedge \star \dd \zeta'_{2} \\
&\qquad \qquad + \frac{2i}{\Delta_{2}} (1+\alpha) \cos[2](\frac{\eta}{2})\dd \zeta'_{1} \wedge \dd \zeta'_{2} \,,
\end{aligned}
\end{equation}
where 
\begin{equation}
\Delta_{2} = 1 + \alpha(2 + \alpha) \cos[2](\frac{\eta}{2})\,.
\end{equation}
The equations of motion for $\zeta'_{1}$ and $\zeta'_{2}$ are written as the conservation laws associated with the $U(1)^{2}$ isometries along $k_{1} = \partial_{\zeta'_{1}}$ and $k_{2} = \partial_{\zeta'_{2}}$. The corresponding Noether currents are given by 
\begin{equation}
\begin{aligned}
J_{1}(\alpha) &= \frac{1}{\Delta_{2}}\left[\sin[2](\frac{\eta}{2}) \dd \zeta'_{1} + i(1 + \alpha) \cos[2](\frac{\eta}{2}) \star \dd \zeta'_{2}\right]\,,\\
J_{2}(\alpha) &= \frac{1}{\Delta_{2}}\left[ \cos[2](\frac{\eta}{2})\dd \zeta'_{2} -  i(1+\alpha)\cos[2](\frac{\eta}{2}) \star \dd \zeta'_{1}\right]\,,
\end{aligned}
\end{equation}
respectively. Up to linear order $\mathcal{O}(\alpha)$, the action~\eqref{eq:TsT-action} is given by 
\begin{equation}
\begin{aligned}
S'_{\alpha} &\sim S_{0}[\eta, \zeta'_{1}, \zeta'_{2}] + \frac{\alpha}{2} \int -2 \sin[2](\frac{\eta}{2})\cos[2](\frac{\eta}{2}) \dd \zeta'_{1} \wedge \star \dd \zeta'_{1} - 2 \cos[4](\frac{\eta}{2}) \dd \zeta'_{2} \wedge \star \dd \zeta'_{2} \\
&\qquad\qquad\qquad\qquad\qquad +2 i \cos[2](\frac{\eta}{2})\sin[2](\frac{\eta}{2}) \dd \zeta'_{1} \wedge \dd \zeta'_{2}\,,
\end{aligned}
\end{equation}
which implies that the contribution of the deformation cannot not be written only in terms of $J_{1}(0)$ and $J_{2}(0)$ in a closed form. 

Applying the $O(d,d)$ map given in~\eqref{eq:Odd-inverse}, we obtain%
\begin{equation}
\begin{aligned}
\dd \zeta_{1} &= \dd \zeta'_{1} +\alpha \star J_{2}(\alpha)\,,\\
\dd \zeta_{2} &=  \dd \zeta'_{2} - \alpha \star J_{1}(\alpha)\,,
\end{aligned}
\end{equation}
which motivates us to define using $J_{\pm} = J_{1} \pm J_{2}$ 
\begin{equation}
\dd \zeta_{\pm} = \dd \zeta'_{\pm} \mp \alpha \star J_{\mp}(\alpha)\,.
\end{equation}
Now that we have identified the transformation of interest, we find the TsT-deformed Lax pairs as follows:
\begin{equation}
\begin{aligned}
\Lax'^{1}_{\LL} =&{}  \biggl( (i b - a ) + (i a - b) \star \biggr) \dd \eta\,,\\
\Lax'^{2}_{\LL} ={}&  -\biggl( (i b - a ) + (i a - b) \star \biggr) \biggl[ \dd \zeta'_{-} - \cos(\eta) \dd \zeta'_{+} + \alpha \star \left(J_{+}(\alpha) + \cos(\eta) J_{-}(\alpha) \right)\biggr] \,,\\
&- \left( \dd \zeta'_{-} +\alpha \star J_{+}(\alpha) \right)\,,\\
\Lax'^{3}_{\LL} ={}&  \biggl( (i  b - a ) + (i a - b) \star \biggr) \sin(\eta)\biggl[ \dd \zeta'_{+} - \alpha \star J_{-}(\alpha)\biggr]\,,
\end{aligned}
\end{equation}
whereas
\begin{equation}\begin{aligned}
\Lax'^{1}_{\RR} ={}& \biggl( (i b + a ) + (i a + b) \star \biggr)\dd \eta\,,\\
\Lax'^{2}_{\RR} ={}& \biggl( (i b + a) + (i a + b)  \star \biggr)\biggl[ \dd \zeta'_{+} -\cos(\eta) \dd \zeta'_{-} - \alpha \star \left( J_{-}(\alpha) + \cos(\eta) J_{+}(\alpha)\right)\biggr] \,,\\
& -\left( \dd \zeta'_{+} - \alpha \star J_{-}(\alpha) \right)\,,\\
\Lax'^{3}_{\RR} ={}&  \biggl( (i b + a) + (i a + b)  \star \biggr) \sin(\eta) \biggl[ \dd \zeta'_{-} + \alpha \star J_{+}(\alpha)\biggr] \,. 
\end{aligned}\end{equation}
We see that the well-known example of the integrable TsT deformation can be also treated in a systematic manner as an $O(d,d)$ transformation.

\subsection{Double T-duality}
As a final explicit example, let us now consider an $O(2,2)$ transformation not connected to the identity that describes a one-parameter deformation of the T-dual model.
The transformation consists of the combination of
\begin{enumerate}
\item an $O(2)\times O(2)$ rotation $g(\alpha)$ given by 
\begin{equation}
g(\alpha) = \pmqty{
	0  & \sin(\alpha) & \cos(\alpha) & 0 \\
	-\sin(\alpha) & 0 & 0 & \cos(\alpha) \\
	\cos(\alpha) & 0 & 0 & \sin(\alpha) \\
	0 & \cos(\alpha) & - \sin(\alpha) & 0 
	}\,.
      \end{equation}
      Note that for \(\alpha = 0\) this \emph{does not} reduce to the identity
\item a diffeomorphism $g_{\mathrm{diff}}$~\eqref{eq:diffeo} characterized by
\begin{equation}
A(\alpha) = \pmqty{
	\cos(\alpha) - \sin(\alpha)  & 0 \\
	0 & \cos(\alpha)
	}\,.
\end{equation}
\item and a $B$-shift $g_{B}$~\eqref{eq:B-shift} given by 
\begin{equation}
\dd \Lambda(\alpha) = 
	\pmqty{
	0 & \sin(\alpha)(\sin(\alpha)+\cos(\alpha)) \\
	- \sin(\alpha)(\sin(\alpha)+\cos(\alpha))  & 0
	}\,.
\end{equation} 
\end{enumerate}
All together, we obtain
\begin{equation}%
  \label{eq:DT-element}
g_{J \bar{J}}(\alpha) =  g(\alpha) g_{\mathrm{diff}}(A(\alpha)) g_{B}(\dd \Lambda(\alpha))
= \pmqty{
	0 & -1 & \cot(\alpha) & 0 \\
	\frac{\tan(\alpha)}{1 - \tan(\alpha)} & 0 & 0 & \frac{1}{1- \tan(\alpha)} \\
	\frac{\tan(\alpha)}{1 - \tan(\alpha)} & 0 & 0 & \frac{\tan(\alpha)}{1 - \tan(\alpha)} \\
	0 & 1 & -1 & 0 
	}.
\end{equation}
Then the generalized metric transforms as
\begin{equation}
\begin{aligned}
  \mathcal{H}^{\prime} &= g_{J \bar{J}}(\alpha)^{t} \mathcal{H}g_{J \bar{J}}(\alpha)\\
  &=
  \pmqty{
   \frac{\tan[2](\alpha)}{\left(1- \tan(\alpha)\right)^{2}} & 0 & 0 & - \frac{\tan[2](\alpha)}{\left(1- \tan(\alpha)\right)^{2}} \\
   0 & \tan[2](\frac{\eta}{2}) & -\tan[2](\frac{\eta}{2}) & 0\\
   0 & -\tan[2](\frac{\eta}{2}) & \frac{\Delta_{2}}{(\sin(\alpha))^{2}\cos[2]( \frac{\eta}{2} ) } & 0 \\
   \frac{\tan[2](\alpha)}{\left(1- \tan(\alpha)\right)^{2}}  & 0 & 0 & \frac{\Delta_{2}}{(\cos(\alpha) - \sin(\alpha))^{2}\sin[2]( \frac{\eta}{2} ) }
   		}\,,
\end{aligned}
\end{equation}
where 
\begin{equation}
\Delta_{3} = \sin[2](\alpha) + \cos[2](\frac{\eta}{2})\cos(\alpha)( \cos(\alpha) - 2\sin(\alpha))\,.
\end{equation}
This leads to the following deformed $\sigma$-model:
\begin{equation}
\label{eq:DT-action}
\begin{aligned}
S'_{\alpha}[\eta, \zeta'_{1}, \zeta'_{2}] ={}& \frac{1}{2}\int \frac{1}{4} \dd \eta \wedge \star \dd \eta + \frac{1}{\Delta_{3}}\sin[2](\alpha) \cos[2](\frac{\eta}{2})\dd \zeta'_{1} \wedge \star \dd \zeta'_{1}\\
& + \frac{1}{\Delta_{3}}(\cos(\alpha) - \sin(\alpha))^{2}\sin[2]( \frac{\eta}{2}) \dd \zeta'_{2} \wedge \star \dd \zeta'_{2} \\
&+ \frac{2i}{\Delta_{3}} \sin[2](\alpha)\sin[2]( \frac{\eta}{2} ) \dd \zeta'_{1} \wedge \dd \zeta'_{2}\,.
\end{aligned}
\end{equation}
Since this $O(2,2)$ element is not in the component connected to the identity, the above deformed action does not turn into the one for the three-sphere for any value of \(\alpha\).

An interesting limit is the $SU(2)/U(1)$ gauged \ac{wznw} model given by $\alpha = 0$~\cite{PhysRevD.44.314,Bardacki:1990wj,Tseytlin:1992ri,Hoare:2014oua},
\begin{equation}
\lim_{\alpha \rightarrow 0}S'_{\alpha}[\eta, \zeta'_{+}, \zeta'_{-}] = \frac{1}{2} \int \frac{1}{4} \dd \eta \wedge \star \dd \eta + \frac{1}{4}\tan[2](\frac{\eta}{2}) \dd (\zeta'_{+} - \zeta'_{-}) \wedge \star \dd (\zeta'_{+} - \zeta'_{-})\,,
\end{equation}
where the metric becomes degenerate and the $B$-field vanishes. 

As before, using~\eqref{eq:DT-action}, the \ac{eom} for $\zeta'_{1,}\,, \zeta'_{2}$ can be expressed as
\begin{align}
\dd \star J_{1}(\alpha) &= 0 , & \dd \star J_{2}(\alpha) = 0,
\end{align}
where we define the conserved currents of the dual model to be
\begin{align}
J_{1}(\alpha) &= \frac{1}{\Delta_{3}}\sin[2](\alpha)\left[ \cos[2](\frac{\eta}{2}) \dd \zeta'_{1} + i \sin[2](\frac{\eta}{2}) \star \dd \zeta'_{2}\right]\,,\\
J_{2}(\alpha) &= \frac{1}{\Delta_{3}}\bqty{\pqty{\cos(\alpha) - \sin(\alpha)}^{2}\sin[2](\frac{\eta}{2})\dd \zeta'_{2} - i \sin[2](\alpha) \sin[2](\frac{\eta}{2}) \star \dd \zeta'_{1} } .
\end{align}
They correspond to the $U(1) \times U(1)$ isometries along the Killing vectors $\partial_{\zeta'_{1}}$ and $\partial_{\zeta'_{1}}$\,, respectively. Note that 
\begin{equation}
S'_{\alpha}[\eta, \zeta'_{1}, \zeta'_{2}] = \frac{1}{2} \int \frac{1}{4} \dd \eta \wedge \star \dd \eta + \dd \zeta'_{1} \wedge \star J_{1}(\alpha) + \dd \zeta'_{2} \wedge \star J_{2}(\alpha)\,.
\end{equation}
Applying the inverse of the $O(d,d)$-duality map in~\eqref{eq:Odd-inverse}, we obtain
\begin{equation}
\begin{aligned}
\dd \zeta_{1} &= - \dd \zeta'_{2} + \cot(\alpha) \star J_{1}(\alpha)\,,\\
\dd \zeta_{2} &=  \frac{\tan(\alpha)}{1 - \tan(\alpha)}\dd \zeta'_{1} - \frac{1}{1- \tan(\alpha)} \star J_{2}(\alpha)\,.
\end{aligned}
\end{equation}
As a result, we find the deformed flat currents $\Lax' = \hat{\Lax}(\dd \zeta_{1/2} \rightarrow \dd \zeta'_{1/2})$:
\begin{equation}\begin{aligned}
\Lax'^{1}_{\LL} ={}& \biggl( (i b - a ) + (i a - b) \star \biggr) \dd \eta\,,\\
\Lax'^{2}_{\LL} ={}& -2\biggl((i b-a) + (i a-b) \star \biggr)
\biggl[ 
	\pmqty{ \sin[2](\frac{\eta}{2}) & \frac{-\cos[2](\frac{\eta}{2})}{1+ \tan(\alpha)} } 
	\pmqty{ \dd \zeta'_{1} + \tan(\alpha) \star J_{2}(\alpha)\\
			  \dd \zeta'_{2} - \tan(\alpha) \star J_{1}(\alpha) }
\biggr]\\
& - \dd \zeta'_{1} - \tan(\alpha) \star J_{2}(\alpha) + \frac{1}{1+\tan(\alpha)}\left( \dd \zeta'_{2} - \tan(\alpha) \star J_{1}(\alpha)\right)\,,\\
\Lax'^{3}_{\LL} ={}& \biggl((i b-a) + (i a-b) \star\biggr) \sin(\eta)\biggl[ \dd \zeta'_{1} + \tan(\alpha) \star J_{2}(\alpha) + \frac{\dd \zeta'_{2} - \tan(\alpha) \star J_{1}(\alpha)}{1+\tan(\alpha)} \biggr] ,
\end{aligned}\end{equation}
whereas
\begin{equation}\begin{aligned}
\Lax'^{1}_{\RR} ={}& \biggl( (i b + a ) + (i a + b) \star \biggr)\dd \eta\,,\\
\Lax'^{2}_{\RR} ={}& 2\biggl( (i b + a) + (i a + b)  \star \biggr)
\biggl[ 
	\pmqty{ \sin[2](\frac{\eta}{2}) & \frac{+\cos[2](\frac{\eta}{2})}{1+ \tan(\alpha)} } 
	\pmqty{ \dd \zeta'_{1} + \tan(\alpha) \star J_{2}(\alpha)\\
			  \dd \zeta'_{2} - \tan(\alpha) \star J_{1}(\alpha) }
\biggr]\\
& - \dd \zeta'_{1} - \tan(\alpha) \star J_{2}(\alpha) - \frac{1}{1+\tan(\alpha)}\left( \dd \zeta'_{2} - \tan(\alpha) \star J_{1}(\alpha)\right)\,,\\
\Lax'^{3}_{\RR} ={}&  \biggl( (i b + a) + (i a + b)  \star \biggr) \sin(\eta) \biggl[ \dd \zeta'_{1} + \tan(\alpha) \star J_{2}(\alpha) - \frac{\dd \zeta'_{2} - \tan(\alpha) \star J_{1}(\alpha)}{1+\tan(\alpha)}\biggr] .
\end{aligned}\end{equation}
Also this example is therefore systematically classified as classically integrable.

\section{Conclusions and Outlook}%
\label{sec:conclusions}
In this paper we showed the classical integrability of models which are obtained from integrable sigma models with a non-Abelian isometry group via an $O(d,d)$ transformation. This class includes T-dual models, $J\bar J$-deformed models and TsT-deformed models.
To show the classical integrability, we have explicitly constructed the transformed Lax pairs using the doubled formalism and shown that they remain flat. 

Our conceptual basis is the string-theoretic picture of the equivalence between momenta and winding modes, showing that integrability goes beyond Noether's construction which is based on the study of isometries only. Locality and non-locality are not physical properties but dependent on the duality frame. 
Our discussion is another example of a case in which the stringy picture is superior to a purely field-theoretic point of view. 

\medskip
Starting from here, it is possible to develop this line of thought in a variety of directions: 
\begin{itemize}
	\item As further concrete examples, one could study \ac{wznw} models with a compact group different from $SU(2)$. This case will yield a picture which is qualitatively similar to the explicit examples discussed here. Since the rank is in general larger than one, the possible transformations are richer and will lead to more interesting deformations.
	\item When studying \ac{wznw} models on non-compact groups, there are several inequivalent maximal tori that can be considered (because they are not related by inner automorphisms), leading in turn to systems with very different physical properties.
	\item Of course \ac{wznw} models are just a natural starting point. We can study in the same way any integrable sigma model with a non-Abelian isometry group of the target space.
	\item $O(d,d)$ transformations capture a number of deformations which can be understood as Yang--Baxter deformations~\cite{Sakamoto:2017cpu,Sakamoto:2018krs}. It would be interesting to understand the extent to which these two approaches are related. A recent work exploring this direction is~\cite{Catal-Ozer:2019tmm}.
	\item It would be interesting to uplift our construction of deformed Lax pairs via the $O(d,d)$-duality map~\eqref{eq:duality-map} to the cases of other T-dualities such as non-Abelian T-duality~\cite{delaOssa:1992vci} and Poisson-Lie T-duality~\cite{Klimcik:1995ux,VonUnge:2002xjf} based on the so-called $\mathcal{E}$-model~\cite{Klimcik:2015gba,Klimcik:2017ken}. In particular, the Poisson-Lie T-duality emerges naturally in Double Field Theory~\cite{Hull:2009mi} as pointed out in~\cite{Hassler:2017yza}. For a related work, see~\cite{Demulder:2018lmj}.
	\item We have explicitly constructed the deformed Lax pairs and have outlined the construction of the corresponding conserved charges in Section~\ref{sec:Odd-currents}. It would be interesting to understand the role of the deformation parameters of the group $O(d,d)$ in the algebra of conserved charges. The method in~\cite{Kluson:2008nn,Kawaguchi:2010jg,Kawaguchi:2011mz,Kawaguchi:2013gma,Delduc:2014uaa} would be helpful.
	\item Since we have focused solely on classical integrability, our results apply both to $O(d,d;\setZ)$ and $O(d,d;\setR)$. The first case amounts to an exact equivalence of models, while $O(d,d;\setR)$ should be understood as a solution-generating technique. This difference should appear when studying the algebra satisfied by the charges. In the latter case, we would expect a deformation of the algebra to appear~\cite{Kawaguchi:2011pf,Kawaguchi:2011wt,Kawaguchi:2012ve,Kawaguchi:2012ug,Kawaguchi:2013lba,Kawaguchi:2013gma}.
	\item In the study of integrability, the spectral parameter $\lambda$ plays an important role. It would be interesting to see whether $O(d,d)$ transformations can be understood in terms of a map acting on the spectral parameter~\cite{Kawaguchi:2012gp}.
        \item One interesting open question that remains is the behavior of integrable models under non-perturbative string dualities such as S-duality.
Since the dual system is supposed to describe the same physics as the original one, a reasonable conjecture is that integrability should be preserved.
The issue with non-perturbative dualities is that at least some of the perturbative local degrees of freedom in one frame become in general non-local and non-perturbative in the other, so it can be very difficult to identify those that are required to realize integrability from the point of view of the two-dimensional theory.
This might explain existing examples in the literature of the non-preservation of integrability in S-dual models~\cite{Giataganas:2013dha}.
\end{itemize}
These points go beyond the scope of this work, but would be interesting topics for future research.

\subsection*{Acknowledgments}

It is our pleasure to thank Chris Hull for discussions and comments on the manuscript and Dan Israel and Junichi Sakamoto for fruitful discussions. Y.S. is grateful to Jean-Pierre Derendinger for instructive discussions, and to the University of Turin for hospitality.
D.O. and S.R. thank the Department of Physics of Kyoto University for hospitality.  D.O. and  Y.S. also thank the Galileo Galilei Institute for Theoretical Physics and INFN for hospitality and partial support during the workshop ``String Theory from a worldsheet perspective'' where part of this work has been done. 
The work of S.R. and Y.S. is supported by the Swiss National Science Foundation (\textsc{snf}) under grant number \textsc{pp00p2\_183718/1}.
D.O.~acknowledges partial support by the NCCR 51NF40--141869 ``The Mathematics of Physics'' (SwissMAP). 
The work of K.Y. is supported by the Supporting Program for Interaction based Initiative Team Studies 
(SPIRITS) from Kyoto University and by a JSPS Grant-in-Aid for Scientific Research (B) No.\,18H01214.   
This work is also supported in part by the JSPS Japan-Russia Research Cooperative Program.

\def\baselinestretch{1}

\begin{small}
  \bibliography{OddRefs}{}

\providecommand{\href}[2]{#2}\begingroup\raggedright\begin{thebibliography}{100}

\bibitem{Hassan:1992gi}
S.~F. Hassan and A.~Sen, \emph{{Marginal deformations of WZNW and coset models
  from O(d,d) transformation}},
  \href{https://doi.org/10.1016/0550-3213(93)90429-S}{\emph{Nucl. Phys.}
  {\bfseries B405} (1993) 143--165},
  [\href{https://arxiv.org/abs/hep-th/9210121}{{\ttfamily hep-th/9210121}}].

\bibitem{Henningson:1992rn}
M.~Henningson and C.~R. Nappi, \emph{{Duality, marginal perturbations and
  gauging}}, \href{https://doi.org/10.1103/PhysRevD.48.861}{\emph{Phys. Rev.}
  {\bfseries D48} (1993) 861--868},
  [\href{https://arxiv.org/abs/hep-th/9301005}{{\ttfamily hep-th/9301005}}].

\bibitem{Kiritsis:1993ju}
E.~Kiritsis, \emph{{Exact duality symmetries in CFT and string theory}},
  \href{https://doi.org/10.1016/0550-3213(93)90428-R}{\emph{Nucl. Phys.}
  {\bfseries B405} (1993) 109--142},
  [\href{https://arxiv.org/abs/hep-th/9302033}{{\ttfamily hep-th/9302033}}].

\bibitem{Giveon:1993ph}
A.~Giveon and E.~Kiritsis, \emph{{Axial vector duality as a gauge symmetry and
  topology change in string theory}},
  \href{https://doi.org/10.1016/0550-3213(94)90460-X}{\emph{Nucl. Phys.}
  {\bfseries B411} (1994) 487--508},
  [\href{https://arxiv.org/abs/hep-th/9303016}{{\ttfamily hep-th/9303016}}].

\bibitem{Forste:1994wp}
S.~Forste, \emph{{A Truly marginal deformation of SL(2, R) in a null
  direction}}, \href{https://doi.org/10.1016/0370-2693(94)91340-4}{\emph{Phys.
  Lett.} {\bfseries B338} (1994) 36--39},
  [\href{https://arxiv.org/abs/hep-th/9407198}{{\ttfamily hep-th/9407198}}].

\bibitem{Israel:2003ry}
D.~Israel, C.~Kounnas and M.~P. Petropoulos, \emph{{Superstrings on NS5
  backgrounds, deformed AdS$_3$ and holography}},
  \href{https://doi.org/10.1088/1126-6708/2003/10/028}{\emph{JHEP} {\bfseries
  10} (2003) 028}, [\href{https://arxiv.org/abs/hep-th/0306053}{{\ttfamily
  hep-th/0306053}}].

\bibitem{Forste:2003km}
S.~Forste and D.~Roggenkamp, \emph{{Current current deformations of conformal
  field theories, and WZW models}},
  \href{https://doi.org/10.1088/1126-6708/2003/05/071}{\emph{JHEP} {\bfseries
  05} (2003) 071}, [\href{https://arxiv.org/abs/hep-th/0304234}{{\ttfamily
  hep-th/0304234}}].

\bibitem{McGough:2016lol}
L.~McGough, M.~Mezei and H.~Verlinde, \emph{{Moving the CFT into the bulk with
  $ T\overline{T} $}},
  \href{https://doi.org/10.1007/JHEP04(2018)010}{\emph{JHEP} {\bfseries 04}
  (2018) 010}, [\href{https://arxiv.org/abs/1611.03470}{{\ttfamily
  1611.03470}}].

\bibitem{Apolo:2018qpq}
L.~Apolo and W.~Song, \emph{{Strings on warped AdS$_{3}$ via $
  \mathrm{T}\bar{\mathrm{J}} $ deformations}},
  \href{https://doi.org/10.1007/JHEP10(2018)165}{\emph{JHEP} {\bfseries 10}
  (2018) 165}, [\href{https://arxiv.org/abs/1806.10127}{{\ttfamily
  1806.10127}}].

\bibitem{Giveon:2017nie}
A.~Giveon, N.~Itzhaki and D.~Kutasov, \emph{{$ \mathrm{T}\overline{\mathrm{T}}
  $ and LST}}, \href{https://doi.org/10.1007/JHEP07(2017)122}{\emph{JHEP}
  {\bfseries 07} (2017) 122},
  [\href{https://arxiv.org/abs/1701.05576}{{\ttfamily 1701.05576}}].

\bibitem{Chakraborty:2018vja}
S.~Chakraborty, A.~Giveon and D.~Kutasov, \emph{{$ J\overline{T} $ deformed
  CFT$_{2}$ and string theory}},
  \href{https://doi.org/10.1007/JHEP10(2018)057}{\emph{JHEP} {\bfseries 10}
  (2018) 057}, [\href{https://arxiv.org/abs/1806.09667}{{\ttfamily
  1806.09667}}].

\bibitem{Borsato:2018spz}
R.~Borsato and L.~Wulff, \emph{{Marginal deformations of WZW models and the
  classical Yang-Baxter equation}},
  \href{https://doi.org/10.1088/1751-8121/ab1b9c}{\emph{J. Phys.} {\bfseries
  A52} (2019) 225401}, [\href{https://arxiv.org/abs/1812.07287}{{\ttfamily
  1812.07287}}].

\bibitem{Araujo:2018rho}
T.~Araujo, E.~Colgain, Y.~Sakatani, M.~M. Sheikh-Jabbari and H.~Yavartanoo,
  \emph{{Holographic integration of $T \bar{T}$ \& $J \bar{T}$ via $O(d,d)$}},
  \href{https://doi.org/10.1007/JHEP03(2019)168}{\emph{JHEP} {\bfseries 03}
  (2019) 168}, [\href{https://arxiv.org/abs/1811.03050}{{\ttfamily
  1811.03050}}].

\bibitem{Chakraborty:2019mdf}
S.~Chakraborty, A.~Giveon and D.~Kutasov, \emph{{$T\bar{T}$, $J\bar{T}$,
  $T\bar{J}$ and String Theory}},
  \href{https://arxiv.org/abs/1905.00051}{{\ttfamily 1905.00051}}.

\bibitem{Chaudhuri:1988qb}
S.~Chaudhuri and J.~A. Schwartz, \emph{{A Criterion for Integrably Marginal
  Operators}}, \href{https://doi.org/10.1016/0370-2693(89)90393-6}{\emph{Phys.
  Lett.} {\bfseries B219} (1989) 291--296}.

\bibitem{Giveon:1988tt}
A.~Giveon, E.~Rabinovici and G.~Veneziano, \emph{{Duality in String Background
  Space}}, \href{https://doi.org/10.1016/0550-3213(89)90489-6}{\emph{Nucl.
  Phys.} {\bfseries B322} (1989) 167--184}.

\bibitem{Duff:1989tf}
M.~J. Duff, \emph{{Duality Rotations in String Theory}},
  \href{https://doi.org/10.1016/0550-3213(90)90520-N}{\emph{Nucl. Phys.}
  {\bfseries B335} (1990) 610}.

\bibitem{Giveon:1991jj}
A.~Giveon and M.~Rocek, \emph{{Generalized duality in curved string
  backgrounds}},
  \href{https://doi.org/10.1016/0550-3213(92)90518-G}{\emph{Nucl. Phys.}
  {\bfseries B380} (1992) 128--146},
  [\href{https://arxiv.org/abs/hep-th/9112070}{{\ttfamily hep-th/9112070}}].

\bibitem{Gasperini:1991qy}
M.~Gasperini, J.~Maharana and G.~Veneziano, \emph{{From trivial to nontrivial
  conformal string backgrounds via O(d,d) transformations}},
  \href{https://doi.org/10.1016/0370-2693(91)91831-F}{\emph{Phys. Lett.}
  {\bfseries B272} (1991) 277--284}.

\bibitem{Maharana:1992my}
J.~Maharana and J.~H. Schwarz, \emph{{Noncompact symmetries in string theory}},
  \href{https://doi.org/10.1016/0550-3213(93)90387-5}{\emph{Nucl. Phys.}
  {\bfseries B390} (1993) 3--32},
  [\href{https://arxiv.org/abs/hep-th/9207016}{{\ttfamily hep-th/9207016}}].

\bibitem{Israel:2004vv}
D.~Israel, C.~Kounnas, D.~Orlando and P.~M. Petropoulos,
  \emph{{Electric/magnetic deformations of S$^3$ and AdS$_3$, and geometric
  cosets}}, \href{https://doi.org/10.1002/prop.200410190}{\emph{Fortsch. Phys.}
  {\bfseries 53} (2005) 73--104},
  [\href{https://arxiv.org/abs/hep-th/0405213}{{\ttfamily hep-th/0405213}}].

\bibitem{Israel:2004cd}
D.~Israel, C.~Kounnas, D.~Orlando and P.~M. Petropoulos, \emph{{Heterotic
  strings on homogeneous spaces}},
  \href{https://doi.org/10.1002/prop.200510250}{\emph{Fortsch. Phys.}
  {\bfseries 53} (2005) 1030--1071},
  [\href{https://arxiv.org/abs/hep-th/0412220}{{\ttfamily hep-th/0412220}}].

\bibitem{Detournay:2005fz}
S.~Detournay, D.~Orlando, P.~M. Petropoulos and P.~Spindel,
  \emph{{Three-dimensional black holes from deformed anti-de Sitter}},
  \href{https://doi.org/10.1088/1126-6708/2005/07/072}{\emph{JHEP} {\bfseries
  07} (2005) 072}, [\href{https://arxiv.org/abs/hep-th/0504231}{{\ttfamily
  hep-th/0504231}}].

\bibitem{Orlando:2006cc}
D.~Orlando, \emph{{String Theory: Exact solutions, marginal deformations and
  hyperbolic spaces}},
  \href{https://doi.org/10.1002/prop.200610333}{\emph{Fortsch. Phys.}
  {\bfseries 55} (2007) 161--282},
  [\href{https://arxiv.org/abs/hep-th/0610284}{{\ttfamily hep-th/0610284}}].

\bibitem{Rennecke:2014sca}
F.~Rennecke, \emph{{O(d,d)-Duality in String Theory}},
  \href{https://doi.org/10.1007/JHEP10(2014)069}{\emph{JHEP} {\bfseries 10}
  (2014) 69}, [\href{https://arxiv.org/abs/1404.0912}{{\ttfamily 1404.0912}}].

\bibitem{Giveon:1994fu}
A.~Giveon, M.~Porrati and E.~Rabinovici, \emph{{Target space duality in string
  theory}}, \href{https://doi.org/10.1016/0370-1573(94)90070-1}{\emph{Phys.
  Rept.} {\bfseries 244} (1994) 77--202},
  [\href{https://arxiv.org/abs/hep-th/9401139}{{\ttfamily hep-th/9401139}}].

\bibitem{Alvarez:1994dn}
E.~Alvarez, L.~Alvarez-Gaume and Y.~Lozano, \emph{{An Introduction to T duality
  in string theory}},
  \href{https://doi.org/10.1016/0920-5632(95)00429-D}{\emph{Nucl. Phys. Proc.
  Suppl.} {\bfseries 41} (1995) 1--20},
  [\href{https://arxiv.org/abs/hep-th/9410237}{{\ttfamily hep-th/9410237}}].

\bibitem{Maharana:2013uvy}
J.~Maharana, \emph{{The Worldsheet Perspective of T-duality Symmetry in String
  Theory}}, \href{https://doi.org/10.1142/S0217751X13300111}{\emph{Int. J. Mod.
  Phys.} {\bfseries A28} (2013) 1330011},
  [\href{https://arxiv.org/abs/1302.1719}{{\ttfamily 1302.1719}}].

\bibitem{Buscher:1987sk}
T.~H. Buscher, \emph{{A Symmetry of the String Background Field Equations}},
  \href{https://doi.org/10.1016/0370-2693(87)90769-6}{\emph{Phys. Lett.}
  {\bfseries B194} (1987) 59--62}.

\bibitem{Buscher:1987qj}
T.~H. Buscher, \emph{{Path Integral Derivation of Quantum Duality in Nonlinear
  Sigma Models}},
  \href{https://doi.org/10.1016/0370-2693(88)90602-8}{\emph{Phys. Lett.}
  {\bfseries B201} (1988) 466--472}.

\bibitem{Tseytlin:1990nb}
A.~A. Tseytlin, \emph{{Duality Symmetric Formulation of String World Sheet
  Dynamics}}, \href{https://doi.org/10.1016/0370-2693(90)91454-J}{\emph{Phys.
  Lett.} {\bfseries B242} (1990) 163--174}.

\bibitem{Tseytlin:1990va}
A.~A. Tseytlin, \emph{{Duality symmetric closed string theory and interacting
  chiral scalars}},
  \href{https://doi.org/10.1016/0550-3213(91)90266-Z}{\emph{Nucl. Phys.}
  {\bfseries B350} (1991) 395--440}.

\bibitem{Hull:2004in}
C.~M. Hull, \emph{{A Geometry for non-geometric string backgrounds}},
  \href{https://doi.org/10.1088/1126-6708/2005/10/065}{\emph{JHEP} {\bfseries
  10} (2005) 065}, [\href{https://arxiv.org/abs/hep-th/0406102}{{\ttfamily
  hep-th/0406102}}].

\bibitem{Hull:2006va}
C.~M. Hull, \emph{{Doubled Geometry and T-Folds}},
  \href{https://doi.org/10.1088/1126-6708/2007/07/080}{\emph{JHEP} {\bfseries
  07} (2007) 080}, [\href{https://arxiv.org/abs/hep-th/0605149}{{\ttfamily
  hep-th/0605149}}].

\bibitem{Hull:2006qs}
C.~M. Hull, \emph{{Global aspects of T-duality, gauged sigma models and
  T-folds}}, \href{https://doi.org/10.1088/1126-6708/2007/10/057}{\emph{JHEP}
  {\bfseries 10} (2007) 057},
  [\href{https://arxiv.org/abs/hep-th/0604178}{{\ttfamily hep-th/0604178}}].

\bibitem{Hull:2007jy}
C.~M. Hull and R.~A. Reid-Edwards, \emph{{Gauge symmetry, T-duality and doubled
  geometry}}, \href{https://doi.org/10.1088/1126-6708/2008/08/043}{\emph{JHEP}
  {\bfseries 08} (2008) 043},
  [\href{https://arxiv.org/abs/0711.4818}{{\ttfamily 0711.4818}}].

\bibitem{Hull:2009sg}
C.~M. Hull and R.~A. Reid-Edwards, \emph{{Non-geometric backgrounds, doubled
  geometry and generalised T-duality}},
  \href{https://doi.org/10.1088/1126-6708/2009/09/014}{\emph{JHEP} {\bfseries
  09} (2009) 014}, [\href{https://arxiv.org/abs/0902.4032}{{\ttfamily
  0902.4032}}].

\bibitem{Dabholkar:2002sy}
A.~Dabholkar and C.~Hull, \emph{{Duality twists, orbifolds, and fluxes}},
  \href{https://doi.org/10.1088/1126-6708/2003/09/054}{\emph{JHEP} {\bfseries
  09} (2003) 054}, [\href{https://arxiv.org/abs/hep-th/0210209}{{\ttfamily
  hep-th/0210209}}].

\bibitem{Dabholkar:2005ve}
A.~Dabholkar and C.~Hull, \emph{{Generalised T-duality and non-geometric
  backgrounds}},
  \href{https://doi.org/10.1088/1126-6708/2006/05/009}{\emph{JHEP} {\bfseries
  05} (2006) 009}, [\href{https://arxiv.org/abs/hep-th/0512005}{{\ttfamily
  hep-th/0512005}}].

\bibitem{Plauschinn:2018wbo}
E.~Plauschinn, \emph{{Non-geometric backgrounds in string theory}},
  \href{https://doi.org/10.1016/j.physrep.2018.12.002}{\emph{Phys. Rept.}
  {\bfseries 798} (2019) 1--122},
  [\href{https://arxiv.org/abs/1811.11203}{{\ttfamily 1811.11203}}].

\bibitem{Luscher:1977uq}
M.~Luscher, \emph{{Quantum Nonlocal Charges and Absence of Particle Production
  in the Two-Dimensional Nonlinear Sigma Model}},
  \href{https://doi.org/10.1016/0550-3213(78)90211-0}{\emph{Nucl. Phys.}
  {\bfseries B135} (1978) 1--19}.

\bibitem{Luscher:1977rq}
M.~Luscher and K.~Pohlmeyer, \emph{{Scattering of Massless Lumps and Nonlocal
  Charges in the Two-Dimensional Classical Nonlinear Sigma Model}},
  \href{https://doi.org/10.1016/0550-3213(78)90049-4}{\emph{Nucl. Phys.}
  {\bfseries B137} (1978) 46--54}.

\bibitem{Brezin:1979am}
E.~Brezin, C.~Itzykson, J.~Zinn-Justin and J.~B. Zuber, \emph{{Remarks About
  the Existence of Nonlocal Charges in Two-Dimensional Models}},
  \href{https://doi.org/10.1016/0370-2693(79)90263-6}{\emph{Phys. Lett.}
  {\bfseries 82B} (1979) 442--444}.

\bibitem{Ricci:2007eq}
R.~Ricci, A.~A. Tseytlin and M.~Wolf, \emph{{On T-Duality and Integrability for
  Strings on AdS Backgrounds}},
  \href{https://doi.org/10.1088/1126-6708/2007/12/082}{\emph{JHEP} {\bfseries
  12} (2007) 082}, [\href{https://arxiv.org/abs/0711.0707}{{\ttfamily
  0711.0707}}].

\bibitem{Hatsuda:2006ts}
M.~Hatsuda and S.~Mizoguchi, \emph{{Nonlocal charges of T-dual strings}},
  \href{https://doi.org/10.1088/1126-6708/2006/07/029}{\emph{JHEP} {\bfseries
  07} (2006) 029}, [\href{https://arxiv.org/abs/hep-th/0603097}{{\ttfamily
  hep-th/0603097}}].

\bibitem{Kluson:2008nn}
J.~Kluson, \emph{{Algebra of Lax Connection for T-Dual Models}},
  \href{https://doi.org/10.1088/1751-8113/42/28/285401}{\emph{J. Phys.}
  {\bfseries A42} (2009) 285401},
  [\href{https://arxiv.org/abs/0812.4510}{{\ttfamily 0812.4510}}].

\bibitem{Orlando:2010yh}
D.~Orlando, S.~Reffert and L.~I. Uruchurtu, \emph{{Classical Integrability of
  the Squashed Three-sphere, Warped AdS$_3$ and Schroedinger Spacetime via
  T-Duality}}, \href{https://doi.org/10.1088/1751-8113/44/11/115401}{\emph{J.
  Phys.} {\bfseries A44} (2011) 115401},
  [\href{https://arxiv.org/abs/1011.1771}{{\ttfamily 1011.1771}}].

\bibitem{Orlando:2012hu}
D.~Orlando and L.~I. Uruchurtu, \emph{{Integrable Superstrings on the Squashed
  Three-sphere}}, \href{https://doi.org/10.1007/JHEP10(2012)007}{\emph{JHEP}
  {\bfseries 10} (2012) 007},
  [\href{https://arxiv.org/abs/1208.3680}{{\ttfamily 1208.3680}}].

\bibitem{Beisert:2010jr}
N.~Beisert et~al., \emph{{Review of AdS/CFT Integrability: An Overview}},
  \href{https://doi.org/10.1007/s11005-011-0529-2}{\emph{Lett. Math. Phys.}
  {\bfseries 99} (2012) 3--32},
  [\href{https://arxiv.org/abs/1012.3982}{{\ttfamily 1012.3982}}].

\bibitem{Maldacena:1997re}
J.~M. Maldacena, \emph{{The Large N limit of superconformal field theories and
  supergravity}}, \href{https://doi.org/10.1023/A:1026654312961,
  10.4310/ATMP.1998.v2.n2.a1}{\emph{Int. J. Theor. Phys.} {\bfseries 38} (1999)
  1113--1133}, [\href{https://arxiv.org/abs/hep-th/9711200}{{\ttfamily
  hep-th/9711200}}].

\bibitem{Bena:2003wd}
I.~Bena, J.~Polchinski and R.~Roiban, \emph{{Hidden symmetries of the
  AdS$_5\times$S$^5$ superstring}},
  \href{https://doi.org/10.1103/PhysRevD.69.046002}{\emph{Phys. Rev.}
  {\bfseries D69} (2004) 046002},
  [\href{https://arxiv.org/abs/hep-th/0305116}{{\ttfamily hep-th/0305116}}].

\bibitem{Alday:2003zb}
L.~F. Alday, \emph{{Nonlocal charges on AdS$_5\times$S$^5$ and PP waves}},
  \href{https://doi.org/10.1088/1126-6708/2003/12/033}{\emph{JHEP} {\bfseries
  12} (2003) 033}, [\href{https://arxiv.org/abs/hep-th/0310146}{{\ttfamily
  hep-th/0310146}}].

\bibitem{Hatsuda:2004it}
M.~Hatsuda and K.~Yoshida, \emph{{Classical integrability and super Yangian of
  superstring on AdS$_5\times$S$^5$}},
  \href{https://doi.org/10.4310/ATMP.2005.v9.n5.a2}{\emph{Adv. Theor. Math.
  Phys.} {\bfseries 9} (2005) 703--728},
  [\href{https://arxiv.org/abs/hep-th/0407044}{{\ttfamily hep-th/0407044}}].

\bibitem{Beisert:2008iq}
N.~Beisert, R.~Ricci, A.~A. Tseytlin and M.~Wolf, \emph{{Dual Superconformal
  Symmetry from AdS$_5\times$S$^5$ Superstring Integrability}},
  \href{https://doi.org/10.1103/PhysRevD.78.126004}{\emph{Phys. Rev.}
  {\bfseries D78} (2008) 126004},
  [\href{https://arxiv.org/abs/0807.3228}{{\ttfamily 0807.3228}}].

\bibitem{Beisert:2009cs}
N.~Beisert, \emph{{T-Duality, Dual Conformal Symmetry and Integrability for
  Strings on AdS$_5\times$S$^5$}},
  \href{https://doi.org/10.1002/prop.200900060}{\emph{Fortsch. Phys.}
  {\bfseries 57} (2009) 329--337},
  [\href{https://arxiv.org/abs/0903.0609}{{\ttfamily 0903.0609}}].

\bibitem{Hatsuda:2011mt}
M.~Hatsuda and K.~Yoshida, \emph{{Super Yangian of superstring on
  AdS$_5\times$S$^5$ revisited}},
  \href{https://doi.org/10.4310/ATMP.2011.v15.n5.a6}{\emph{Adv. Theor. Math.
  Phys.} {\bfseries 15} (2011) 1485--1501},
  [\href{https://arxiv.org/abs/1107.4673}{{\ttfamily 1107.4673}}].

\bibitem{Klimcik:2002zj}
C.~Klimcik, \emph{{Yang-Baxter sigma models and dS/AdS T duality}},
  \href{https://doi.org/10.1088/1126-6708/2002/12/051}{\emph{JHEP} {\bfseries
  12} (2002) 051}, [\href{https://arxiv.org/abs/hep-th/0210095}{{\ttfamily
  hep-th/0210095}}].

\bibitem{Klimcik:2008eq}
C.~Klimcik, \emph{{On integrability of the Yang-Baxter sigma-model}},
  \href{https://doi.org/10.1063/1.3116242}{\emph{J. Math. Phys.} {\bfseries 50}
  (2009) 043508}, [\href{https://arxiv.org/abs/0802.3518}{{\ttfamily
  0802.3518}}].

\bibitem{Delduc:2013qra}
F.~Delduc, M.~Magro and B.~Vicedo, \emph{{An integrable deformation of the
  AdS$_{5}\times$S$^5$ superstring action}},
  \href{https://doi.org/10.1103/PhysRevLett.112.051601}{\emph{Phys. Rev. Lett.}
  {\bfseries 112} (2014) 051601},
  [\href{https://arxiv.org/abs/1309.5850}{{\ttfamily 1309.5850}}].

\bibitem{Kawaguchi:2014qwa}
I.~Kawaguchi, T.~Matsumoto and K.~Yoshida, \emph{{Jordanian deformations of the
  AdS$_5\times$S$^5$ superstring}},
  \href{https://doi.org/10.1007/JHEP04(2014)153}{\emph{JHEP} {\bfseries 04}
  (2014) 153}, [\href{https://arxiv.org/abs/1401.4855}{{\ttfamily 1401.4855}}].

\bibitem{Klimcik:2014bta}
C.~Klimcik, \emph{{Integrability of the bi-Yang-Baxter sigma-model}},
  \href{https://doi.org/10.1007/s11005-014-0709-y}{\emph{Lett. Math. Phys.}
  {\bfseries 104} (2014) 1095--1106},
  [\href{https://arxiv.org/abs/1402.2105}{{\ttfamily 1402.2105}}].

\bibitem{Klimcik:2015gba}
C.~Klimcik, \emph{{Eta and lambda deformations as E-models}},
  \href{https://doi.org/10.1016/j.nuclphysb.2015.09.011}{\emph{Nucl. Phys.}
  {\bfseries B900} (2015) 259--272},
  [\href{https://arxiv.org/abs/1508.05832}{{\ttfamily 1508.05832}}].

\bibitem{Klimcik:2016rov}
C.~Klimcik, \emph{{Poisson--Lie T-duals of the bi-Yang--Baxter models}},
  \href{https://doi.org/10.1016/j.physletb.2016.06.077}{\emph{Phys. Lett.}
  {\bfseries B760} (2016) 345--349},
  [\href{https://arxiv.org/abs/1606.03016}{{\ttfamily 1606.03016}}].

\bibitem{Klimcik:2017ken}
C.~Klimcik, \emph{{Yang-Baxter $\sigma$-model with WZNW term as ${\mathcal
  E}$-model}},
  \href{https://doi.org/10.1016/j.physletb.2017.07.051}{\emph{Phys. Lett.}
  {\bfseries B772} (2017) 725--730},
  [\href{https://arxiv.org/abs/1706.08912}{{\ttfamily 1706.08912}}].

\bibitem{Hashimoto:1999ut}
A.~Hashimoto and N.~Itzhaki, \emph{{Noncommutative Yang-Mills and the AdS/CFT
  correspondence}},
  \href{https://doi.org/10.1016/S0370-2693(99)01037-0}{\emph{Phys. Lett.}
  {\bfseries B465} (1999) 142--147},
  [\href{https://arxiv.org/abs/hep-th/9907166}{{\ttfamily hep-th/9907166}}].

\bibitem{Maldacena:1999mh}
J.~M. Maldacena and J.~G. Russo, \emph{{Large N limit of noncommutative gauge
  theories}}, \href{https://doi.org/10.1088/1126-6708/1999/09/025}{\emph{JHEP}
  {\bfseries 09} (1999) 025},
  [\href{https://arxiv.org/abs/hep-th/9908134}{{\ttfamily hep-th/9908134}}].

\bibitem{Lunin:2005jy}
O.~Lunin and J.~M. Maldacena, \emph{{Deforming field theories with
  U(1)$\times$U(1) global symmetry and their gravity duals}},
  \href{https://doi.org/10.1088/1126-6708/2005/05/033}{\emph{JHEP} {\bfseries
  05} (2005) 033}, [\href{https://arxiv.org/abs/hep-th/0502086}{{\ttfamily
  hep-th/0502086}}].

\bibitem{Alday:2005ww}
L.~F. Alday, G.~Arutyunov and S.~Frolov, \emph{{Green-Schwarz strings in
  TsT-transformed backgrounds}},
  \href{https://doi.org/10.1088/1126-6708/2006/06/018}{\emph{JHEP} {\bfseries
  06} (2006) 018}, [\href{https://arxiv.org/abs/hep-th/0512253}{{\ttfamily
  hep-th/0512253}}].

\bibitem{Frolov:2005dj}
S.~Frolov, \emph{{Lax pair for strings in Lunin-Maldacena background}},
  \href{https://doi.org/10.1088/1126-6708/2005/05/069}{\emph{JHEP} {\bfseries
  05} (2005) 069}, [\href{https://arxiv.org/abs/hep-th/0503201}{{\ttfamily
  hep-th/0503201}}].

\bibitem{Matsumoto:2014ubv}
T.~Matsumoto and K.~Yoshida, \emph{{Yang-Baxter deformations and string
  dualities}}, \href{https://doi.org/10.1007/JHEP03(2015)137}{\emph{JHEP}
  {\bfseries 03} (2015) 137},
  [\href{https://arxiv.org/abs/1412.3658}{{\ttfamily 1412.3658}}].

\bibitem{Matsumoto:2014nra}
T.~Matsumoto and K.~Yoshida, \emph{{Lunin-Maldacena backgrounds from the
  classical Yang-Baxter equation - towards the gravity/CYBE correspondence}},
  \href{https://doi.org/10.1007/JHEP06(2014)135}{\emph{JHEP} {\bfseries 06}
  (2014) 135}, [\href{https://arxiv.org/abs/1404.1838}{{\ttfamily 1404.1838}}].

\bibitem{Crichigno:2014ipa}
P.~M. Crichigno, T.~Matsumoto and K.~Yoshida, \emph{{Deformations of $T^{1,1}$
  as Yang-Baxter sigma models}},
  \href{https://doi.org/10.1007/JHEP12(2014)085}{\emph{JHEP} {\bfseries 12}
  (2014) 085}, [\href{https://arxiv.org/abs/1406.2249}{{\ttfamily 1406.2249}}].

\bibitem{Matsumoto:2014gwa}
T.~Matsumoto and K.~Yoshida, \emph{{Integrability of classical strings dual for
  noncommutative gauge theories}},
  \href{https://doi.org/10.1007/JHEP06(2014)163}{\emph{JHEP} {\bfseries 06}
  (2014) 163}, [\href{https://arxiv.org/abs/1404.3657}{{\ttfamily 1404.3657}}].

\bibitem{Matsumoto:2015ypa}
T.~Matsumoto, D.~Orlando, S.~Reffert, J.~Sakamoto and K.~Yoshida,
  \emph{{Yang-Baxter deformations of Minkowski spacetime}},
  \href{https://doi.org/10.1007/JHEP10(2015)185}{\emph{JHEP} {\bfseries 10}
  (2015) 185}, [\href{https://arxiv.org/abs/1505.04553}{{\ttfamily
  1505.04553}}].

\bibitem{Matsumoto:2015uja}
T.~Matsumoto and K.~Yoshida, \emph{{Schroedinger geometries arising from
  Yang-Baxter deformations}},
  \href{https://doi.org/10.1007/JHEP04(2015)180}{\emph{JHEP} {\bfseries 04}
  (2015) 180}, [\href{https://arxiv.org/abs/1502.00740}{{\ttfamily
  1502.00740}}].

\bibitem{vanTongeren:2015soa}
S.~J. van Tongeren, \emph{{On classical Yang-Baxter based deformations of the
  AdS$_5\times$S$^5$ superstring}},
  \href{https://doi.org/10.1007/JHEP06(2015)048}{\emph{JHEP} {\bfseries 06}
  (2015) 048}, [\href{https://arxiv.org/abs/1504.05516}{{\ttfamily
  1504.05516}}].

\bibitem{Osten:2016dvf}
D.~Osten and S.~J. van Tongeren, \emph{{Abelian Yang--Baxter deformations and
  TsT transformations}},
  \href{https://doi.org/10.1016/j.nuclphysb.2016.12.007}{\emph{Nucl. Phys.}
  {\bfseries B915} (2017) 184--205},
  [\href{https://arxiv.org/abs/1608.08504}{{\ttfamily 1608.08504}}].

\bibitem{Delduc:2017fib}
F.~Delduc, B.~Hoare, T.~Kameyama and M.~Magro, \emph{{Combining the
  bi-Yang-Baxter deformation, the Wess-Zumino term and TsT transformations in
  one integrable $\sigma$-model}},
  \href{https://doi.org/10.1007/JHEP10(2017)212}{\emph{JHEP} {\bfseries 10}
  (2017) 212}, [\href{https://arxiv.org/abs/1707.08371}{{\ttfamily
  1707.08371}}].

\bibitem{Hoare:2016wsk}
B.~Hoare and A.~A. Tseytlin, \emph{{Homogeneous Yang-Baxter deformations as
  non-abelian duals of the $AdS_5$ sigma-model}},
  \href{https://doi.org/10.1088/1751-8113/49/49/494001}{\emph{J. Phys.}
  {\bfseries A49} (2016) 494001},
  [\href{https://arxiv.org/abs/1609.02550}{{\ttfamily 1609.02550}}].

\bibitem{Borsato:2016pas}
R.~Borsato and L.~Wulff, \emph{{Integrable Deformations of $T$-Dual $\sigma$
  Models}}, \href{https://doi.org/10.1103/PhysRevLett.117.251602}{\emph{Phys.
  Rev. Lett.} {\bfseries 117} (2016) 251602},
  [\href{https://arxiv.org/abs/1609.09834}{{\ttfamily 1609.09834}}].

\bibitem{Sakamoto:2016ppx}
J.~Sakamoto and K.~Yoshida, \emph{{Yang-Baxter deformations of $W_{2,4}\times
  T^{1,1}$ and the associated T-dual models}},
  \href{https://doi.org/10.1016/j.nuclphysb.2017.06.017}{\emph{Nucl. Phys.}
  {\bfseries B921} (2017) 805--828},
  [\href{https://arxiv.org/abs/1612.08615}{{\ttfamily 1612.08615}}].

\bibitem{Borsato:2018idb}
R.~Borsato and L.~Wulff, \emph{{Non-abelian T-duality and Yang-Baxter
  deformations of Green-Schwarz strings}},
  \href{https://doi.org/10.1007/JHEP08(2018)027}{\emph{JHEP} {\bfseries 08}
  (2018) 027}, [\href{https://arxiv.org/abs/1806.04083}{{\ttfamily
  1806.04083}}].

\bibitem{Lust:2018jsx}
D.~Luest and D.~Osten, \emph{{Generalised fluxes, Yang-Baxter deformations and
  the O(d,d) structure of non-abelian T-duality}},
  \href{https://doi.org/10.1007/JHEP05(2018)165}{\emph{JHEP} {\bfseries 05}
  (2018) 165}, [\href{https://arxiv.org/abs/1803.03971}{{\ttfamily
  1803.03971}}].

\bibitem{Arutyunov:2015mqj}
G.~Arutyunov, S.~Frolov, B.~Hoare, R.~Roiban and A.~A. Tseytlin, \emph{{Scale
  invariance of the $\eta$-deformed $AdS_5\times S^5$ superstring, T-duality
  and modified type II equations}},
  \href{https://doi.org/10.1016/j.nuclphysb.2015.12.012}{\emph{Nucl. Phys.}
  {\bfseries B903} (2016) 262--303},
  [\href{https://arxiv.org/abs/1511.05795}{{\ttfamily 1511.05795}}].

\bibitem{Orlando:2016qqu}
D.~Orlando, S.~Reffert, J.~Sakamoto and K.~Yoshida, \emph{{Generalized type IIB
  supergravity equations and non-Abelian classical r-matrices}},
  \href{https://doi.org/10.1088/1751-8113/49/44/445403}{\emph{J. Phys.}
  {\bfseries A49} (2016) 445403},
  [\href{https://arxiv.org/abs/1607.00795}{{\ttfamily 1607.00795}}].

\bibitem{Wulff:2016tju}
A.~A. Tseytlin and L.~Wulff, \emph{{Kappa-symmetry of superstring sigma model
  and generalized 10d supergravity equations}},
  \href{https://doi.org/10.1007/JHEP06(2016)174}{\emph{JHEP} {\bfseries 06}
  (2016) 174}, [\href{https://arxiv.org/abs/1605.04884}{{\ttfamily
  1605.04884}}].

\bibitem{Fernandez-Melgarejo:2017oyu}
J.~J. Fernandez-Melgarejo, J.~Sakamoto, Y.~Sakatani and K.~Yoshida,
  \emph{{$T$-folds from Yang-Baxter deformations}},
  \href{https://doi.org/10.1007/JHEP12(2017)108}{\emph{JHEP} {\bfseries 12}
  (2017) 108}, [\href{https://arxiv.org/abs/1710.06849}{{\ttfamily
  1710.06849}}].

\bibitem{Araujo:2017jkb}
T.~Araujo, I.~Bakhmatov, E.~O. Colgain, J.~Sakamoto, M.~M. Sheikh-Jabbari and
  K.~Yoshida, \emph{{Yang-Baxter $\sigma$-models, conformal twists, and
  noncommutative Yang-Mills theory}},
  \href{https://doi.org/10.1103/PhysRevD.95.105006}{\emph{Phys. Rev.}
  {\bfseries D95} (2017) 105006},
  [\href{https://arxiv.org/abs/1702.02861}{{\ttfamily 1702.02861}}].

\bibitem{Araujo:2017enj}
T.~Araujo, E.~O~Colgain, J.~Sakamoto, M.~M. Sheikh-Jabbari and K.~Yoshida,
  \emph{{$I$ in generalized supergravity}},
  \href{https://doi.org/10.1140/epjc/s10052-017-5316-5}{\emph{Eur. Phys. J.}
  {\bfseries C77} (2017) 739},
  [\href{https://arxiv.org/abs/1708.03163}{{\ttfamily 1708.03163}}].

\bibitem{Araujo:2017jap}
T.~Araujo, I.~Bakhmatov, E.~O. Colgain, J.~Sakamoto, M.~M. Sheikh-Jabbari and
  K.~Yoshida, \emph{{Conformal twists, Yang--Baxter $\sigma$-models \&
  holographic noncommutativity}},
  \href{https://doi.org/10.1088/1751-8121/aac195}{\emph{J. Phys.} {\bfseries
  A51} (2018) 235401}, [\href{https://arxiv.org/abs/1705.02063}{{\ttfamily
  1705.02063}}].

\bibitem{Sakatani:2016fvh}
Y.~Sakatani, S.~Uehara and K.~Yoshida, \emph{{Generalized gravity from modified
  DFT}}, \href{https://doi.org/10.1007/JHEP04(2017)123}{\emph{JHEP} {\bfseries
  04} (2017) 123}, [\href{https://arxiv.org/abs/1611.05856}{{\ttfamily
  1611.05856}}].

\bibitem{Sakamoto:2017wor}
J.~Sakamoto, Y.~Sakatani and K.~Yoshida, \emph{{Weyl invariance for generalized
  supergravity backgrounds from the doubled formalism}},
  \href{https://doi.org/10.1093/ptep/ptx067}{\emph{PTEP} {\bfseries 2017}
  (2017) 053B07}, [\href{https://arxiv.org/abs/1703.09213}{{\ttfamily
  1703.09213}}].

\bibitem{Catal-Ozer:2019tmm}
A.~Catal-Ozer and S.~Tunali, \emph{{Yang-Baxter Deformation as an O(d,d)
  Transformation}},  \href{https://arxiv.org/abs/1906.09053}{{\ttfamily
  1906.09053}}.

\bibitem{Polchinski:1998rq}
J.~Polchinski, \emph{{String theory. Vol. 1: An introduction to the bosonic
  string}}.
\newblock Cambridge Monographs on Mathematical Physics. Cambridge University
  Press, 2007,
  \href{https://doi.org/10.1017/CBO9780511816079}{10.1017/CBO9780511816079}.

\bibitem{Seiberg:1999vs}
N.~Seiberg and E.~Witten, \emph{{String theory and noncommutative geometry}},
  \href{https://doi.org/10.1088/1126-6708/1999/09/032}{\emph{JHEP} {\bfseries
  09} (1999) 032}, [\href{https://arxiv.org/abs/hep-th/9908142}{{\ttfamily
  hep-th/9908142}}].

\bibitem{Andriot:2013xca}
D.~Andriot and A.~Betz, \emph{{$\beta$-supergravity: a ten-dimensional theory
  with non-geometric fluxes, and its geometric framework}},
  \href{https://doi.org/10.1007/JHEP12(2013)083}{\emph{JHEP} {\bfseries 12}
  (2013) 083}, [\href{https://arxiv.org/abs/1306.4381}{{\ttfamily 1306.4381}}].

\bibitem{Blumenhagen:2013aia}
R.~Blumenhagen, A.~Deser, E.~Plauschinn, F.~Rennecke and C.~Schmid, \emph{{The
  Intriguing Structure of Non-geometric Frames in String Theory}},
  \href{https://doi.org/10.1002/prop.201300013}{\emph{Fortsch. Phys.}
  {\bfseries 61} (2013) 893--925},
  [\href{https://arxiv.org/abs/1304.2784}{{\ttfamily 1304.2784}}].

\bibitem{Plauschinn:2014nha}
E.~Plauschinn, \emph{{On T-duality transformations for the three-sphere}},
  \href{https://doi.org/10.1016/j.nuclphysb.2015.02.008}{\emph{Nucl. Phys.}
  {\bfseries B893} (2015) 257--286},
  [\href{https://arxiv.org/abs/1408.1715}{{\ttfamily 1408.1715}}].

\bibitem{PhysRevD.44.314}
E.~Witten, \emph{String theory and black holes},
  \href{https://doi.org/10.1103/PhysRevD.44.314}{\emph{Phys. Rev. D} {\bfseries
  44} (Jul, 1991) 314--324}.

\bibitem{Bardacki:1990wj}
K.~Bardakci, M.~J. Crescimanno and E.~Rabinovici, \emph{{Parafermions From
  Coset Models}},
  \href{https://doi.org/10.1016/0550-3213(90)90365-K}{\emph{Nucl. Phys.}
  {\bfseries B344} (1990) 344--370}.

\bibitem{Tseytlin:1992ri}
A.~A. Tseytlin, \emph{{Effective action of gauged WZW model and exact string
  solutions}}, \href{https://doi.org/10.1016/0550-3213(93)90511-M}{\emph{Nucl.
  Phys.} {\bfseries B399} (1993) 601--622},
  [\href{https://arxiv.org/abs/hep-th/9301015}{{\ttfamily hep-th/9301015}}].

\bibitem{Hoare:2014oua}
B.~Hoare, \emph{{Towards a two-parameter q-deformation of AdS$_3 \times S^3
  \times M^4$ superstrings}},
  \href{https://doi.org/10.1016/j.nuclphysb.2014.12.012}{\emph{Nucl. Phys.}
  {\bfseries B891} (2015) 259--295},
  [\href{https://arxiv.org/abs/1411.1266}{{\ttfamily 1411.1266}}].

\bibitem{Sakamoto:2017cpu}
J.~Sakamoto, Y.~Sakatani and K.~Yoshida, \emph{{Homogeneous Yang-Baxter
  deformations as generalized diffeomorphisms}},
  \href{https://doi.org/10.1088/1751-8121/aa8896}{\emph{J. Phys.} {\bfseries
  A50} (2017) 415401}, [\href{https://arxiv.org/abs/1705.07116}{{\ttfamily
  1705.07116}}].

\bibitem{Sakamoto:2018krs}
J.~Sakamoto and Y.~Sakatani, \emph{{Local $\beta$-deformations and Yang-Baxter
  sigma model}}, \href{https://doi.org/10.1007/JHEP06(2018)147}{\emph{JHEP}
  {\bfseries 06} (2018) 147},
  [\href{https://arxiv.org/abs/1803.05903}{{\ttfamily 1803.05903}}].

\bibitem{delaOssa:1992vci}
X.~C. de~la Ossa and F.~Quevedo, \emph{{Duality symmetries from nonAbelian
  isometries in string theory}},
  \href{https://doi.org/10.1016/0550-3213(93)90041-M}{\emph{Nucl. Phys.}
  {\bfseries B403} (1993) 377--394},
  [\href{https://arxiv.org/abs/hep-th/9210021}{{\ttfamily hep-th/9210021}}].

\bibitem{Klimcik:1995ux}
C.~Klimcik and P.~Severa, \emph{{Dual nonAbelian duality and the Drinfeld
  double}}, \href{https://doi.org/10.1016/0370-2693(95)00451-P}{\emph{Phys.
  Lett.} {\bfseries B351} (1995) 455--462},
  [\href{https://arxiv.org/abs/hep-th/9502122}{{\ttfamily hep-th/9502122}}].

\bibitem{VonUnge:2002xjf}
R.~Von~Unge, \emph{{Poisson Lie T plurality}},
  \href{https://doi.org/10.1088/1126-6708/2002/07/014}{\emph{JHEP} {\bfseries
  07} (2002) 014}, [\href{https://arxiv.org/abs/hep-th/0205245}{{\ttfamily
  hep-th/0205245}}].

\bibitem{Hull:2009mi}
C.~Hull and B.~Zwiebach, \emph{{Double Field Theory}},
  \href{https://doi.org/10.1088/1126-6708/2009/09/099}{\emph{JHEP} {\bfseries
  09} (2009) 099}, [\href{https://arxiv.org/abs/0904.4664}{{\ttfamily
  0904.4664}}].

\bibitem{Hassler:2017yza}
F.~Hassler, \emph{{Poisson-Lie T-Duality in Double Field Theory}},
  \href{https://arxiv.org/abs/1707.08624}{{\ttfamily 1707.08624}}.

\bibitem{Demulder:2018lmj}
S.~Demulder, F.~Hassler and D.~C. Thompson, \emph{{Doubled aspects of
  generalised dualities and integrable deformations}},
  \href{https://doi.org/10.1007/JHEP02(2019)189}{\emph{JHEP} {\bfseries 02}
  (2019) 189}, [\href{https://arxiv.org/abs/1810.11446}{{\ttfamily
  1810.11446}}].

\bibitem{Kawaguchi:2010jg}
I.~Kawaguchi and K.~Yoshida, \emph{{Hidden Yangian symmetry in sigma model on
  squashed sphere}}, \href{https://doi.org/10.1007/JHEP11(2010)032}{\emph{JHEP}
  {\bfseries 11} (2010) 032},
  [\href{https://arxiv.org/abs/1008.0776}{{\ttfamily 1008.0776}}].

\bibitem{Kawaguchi:2011mz}
I.~Kawaguchi, D.~Orlando and K.~Yoshida, \emph{{Yangian symmetry in deformed
  WZNW models on squashed spheres}},
  \href{https://doi.org/10.1016/j.physletb.2011.06.007}{\emph{Phys. Lett.}
  {\bfseries B701} (2011) 475--480},
  [\href{https://arxiv.org/abs/1104.0738}{{\ttfamily 1104.0738}}].

\bibitem{Kawaguchi:2013gma}
I.~Kawaguchi and K.~Yoshida, \emph{{A deformation of quantum affine algebra in
  squashed Wess-Zumino-Novikov-Witten models}},
  \href{https://doi.org/10.1063/1.4880341}{\emph{J. Math. Phys.} {\bfseries 55}
  (2014) 062302}, [\href{https://arxiv.org/abs/1311.4696}{{\ttfamily
  1311.4696}}].

\bibitem{Delduc:2014uaa}
F.~Delduc, M.~Magro and B.~Vicedo, \emph{{Integrable double deformation of the
  principal chiral model}},
  \href{https://doi.org/10.1016/j.nuclphysb.2014.12.018}{\emph{Nucl. Phys.}
  {\bfseries B891} (2015) 312--321},
  [\href{https://arxiv.org/abs/1410.8066}{{\ttfamily 1410.8066}}].

\bibitem{Kawaguchi:2011pf}
I.~Kawaguchi and K.~Yoshida, \emph{{Hybrid classical integrability in squashed
  sigma models}},
  \href{https://doi.org/10.1016/j.physletb.2011.09.117}{\emph{Phys. Lett.}
  {\bfseries B705} (2011) 251--254},
  [\href{https://arxiv.org/abs/1107.3662}{{\ttfamily 1107.3662}}].

\bibitem{Kawaguchi:2011wt}
I.~Kawaguchi and K.~Yoshida, \emph{{Classical integrability of Schrodinger
  sigma models and q-deformed Poincare symmetry}},
  \href{https://doi.org/10.1007/JHEP11(2011)094}{\emph{JHEP} {\bfseries 11}
  (2011) 094}, [\href{https://arxiv.org/abs/1109.0872}{{\ttfamily 1109.0872}}].

\bibitem{Kawaguchi:2012ve}
I.~Kawaguchi, T.~Matsumoto and K.~Yoshida, \emph{{The classical origin of
  quantum affine algebra in squashed sigma models}},
  \href{https://doi.org/10.1007/JHEP04(2012)115}{\emph{JHEP} {\bfseries 04}
  (2012) 115}, [\href{https://arxiv.org/abs/1201.3058}{{\ttfamily 1201.3058}}].

\bibitem{Kawaguchi:2012ug}
I.~Kawaguchi and K.~Yoshida, \emph{{Exotic symmetry and monodromy equivalence
  in Schrodinger sigma models}},
  \href{https://doi.org/10.1007/JHEP02(2013)024}{\emph{JHEP} {\bfseries 02}
  (2013) 024}, [\href{https://arxiv.org/abs/1209.4147}{{\ttfamily 1209.4147}}].

\bibitem{Kawaguchi:2013lba}
I.~Kawaguchi, T.~Matsumoto and K.~Yoshida, \emph{{Schroedinger sigma models and
  Jordanian twists}},
  \href{https://doi.org/10.1007/JHEP08(2013)013}{\emph{JHEP} {\bfseries 08}
  (2013) 013}, [\href{https://arxiv.org/abs/1305.6556}{{\ttfamily 1305.6556}}].

\bibitem{Kawaguchi:2012gp}
I.~Kawaguchi, T.~Matsumoto and K.~Yoshida, \emph{{On the classical equivalence
  of monodromy matrices in squashed sigma model}},
  \href{https://doi.org/10.1007/JHEP06(2012)082}{\emph{JHEP} {\bfseries 06}
  (2012) 082}, [\href{https://arxiv.org/abs/1203.3400}{{\ttfamily 1203.3400}}].

\bibitem{Giataganas:2013dha}
D.~Giataganas, L.~A. Pando~Zayas and K.~Zoubos, \emph{{On Marginal Deformations
  and Non-Integrability}},
  \href{https://doi.org/10.1007/JHEP01(2014)129}{\emph{JHEP} {\bfseries 01}
  (2014) 129}, [\href{https://arxiv.org/abs/1311.3241}{{\ttfamily 1311.3241}}].

\end{thebibliography}\endgroup
  \bibliographystyle{JHEP}
\end{small}

\end{document}